\documentclass[twocolumn]{autart} 

\newtheorem{definition}{Definition}
\newtheorem{lemma}{Lemma}
\newtheorem{remark}{Remark}
\usepackage{bbm}
\usepackage{subfigure}% subcaptions for subfigures
\usepackage{subfigmat}% matrices of similar subfigures, aka small mulitples
\usepackage{amsfonts}
\usepackage{algorithm}
\usepackage{algpseudocode}
\usepackage{amsmath}
\usepackage{amssymb}
\usepackage{graphicx}      % include this line if your document contains figures
\usepackage{xcolor}
\usepackage{comment}
\usepackage{subfigure}
\usepackage{tikz,pgfplots}
\usepackage{multirow}
\pgfplotsset{compat=1.15}

\newcommand{\qedsymbol}{\rule{0.7em}{0.7em}}

\newcommand{\RBFN}{{\rm{RBFN}}}
\newcommand{\q}{w}
\newcommand{\Q}{\mathbb{W}}

\newcommand{\w}{\sigma}
\newcommand{\W}{\mathcal{S}}

\renewcommand{\t}{t}
\newcommand{\gi}{\psi}

\newcommand{\X}{\mathbb{X}}
\renewcommand{\Pr}{\mathsf{Pr}}
\newcommand{\Pq}{\mathsf{Pr}_{\Q}}
\newcommand{\PqN}{\mathsf{Pr}_{\Q^N}}
\newcommand{\Pwb}{\mathsf{Pr}_{\pmb{\sigma}}}
\newcommand{\Su}{{\mathbb{S}}}
\newcommand{\Sug}{\Su(\gamma)}

\newcommand{\scale}{\gamma}

\newcommand{\EE}{\mathbb{E}}

\def\set#1#2{\{ \; #1 \;:\;#2\;\}} % creates a set: { #1 : #2 }

\def\NATS#1{[#1]}

\def\CCS{$\varepsilon$-CCS}
\def\volp{\mathsf{Vol}_p}

\newcommand{\R}{\mathbb{R}}

\newcommand{\Xe}{\mathbb{X}_\varepsilon}

\def\fracg#1#2{{\displaystyle{\frac{#1}{#2}}}}

\newcommand{\Bin}{\mathbf{B}}

\renewcommand{\Pr}{\mathsf{Pr}}
\newcommand{\viol}{\mathsf{Viol}}

\newtheorem{corollary}{Corollary}
\newtheorem {property}{Property}
\newtheorem {example}{Example}
\newtheorem {problem}{Problem}
\newtheorem {theorem}{Theorem}
\usepackage[utf8]{inputenc}

\def\FPS{{\mathsf{FPS}}}

\def\NSp{{\mathbb{S}_p}}
\def\NS1{{\mathbb{S}_1}}
\def\NS2{{\mathbb{S}_2}}

\def\Sum#1#2{\sum\limits_{#1}^{#2}}

\newcommand {\bsis} {\left\{ \begin{array} }
\newcommand {\esis} {\end{array}\right.}
\def\conv#1#2{ \left( \begin{array}{c} #1 \\ #2 \\ \end{array}\right)} 
\def\bmat#1{\left[\begin{array}{#1}}
\def\emat{\end{array}\right]}

\begin{document}

\begin{frontmatter}

\title{Chance constrained sets approximation:\\ A probabilistic scaling 
approach - EXTENDED VERSION}
%\thanks{\small This research was partially funded by the project PRIN 2017, Prot. 2017S559BB, and has received support by Ministerio de Ciencia e Innovación of Spain under project PID2019-106212RB-C41.}

\author[s1]{M. Mammarella}\ead{martina.mammarella@ieiit.cnr.it}
\author[s2]{V. Mirasierra}\ead{vmirasierra@us.es}
\author[s3]{M. Lorenzen}\ead{lorenzen@ist.uni-stuttgart.de}
\author[s2]{T. Alamo}\ead{talamo@us.es}
\author[s1]{F. Dabbene}\ead{fabrizio.dabbene@ieiit.cnr.it}

\address[s1]{CNR-IEIIT; c/o Politecnico di Torino; C.so Duca degli Abruzzi 24, Torino; Italy.}
\address[s2]{Universidad de Sevilla, Escuela Superior de Ingenieros, Camino de los Descubrimientos s/n, Sevilla; Spain.}
\address[s3]{Systemwissenschaften, TTI GmbH, Nobelstr. 15, 70569 Stuttgart, Germany}

\begin{abstract}
In this paper, a sample-based procedure for obtaining simple and computable approximations of chance-contrained sets is proposed. The procedure allows to control the complexity of the approximating set, by defining families of simple-approximating sets of given complexity. A probabilistic scaling procedure then allows to rescale these sets to obtain the desired probabilistic guarantees. The proposed approach is shown to be applicable in several problem in systems and control, such as the design of Stochastic Model Predictive Control schemes or the solution of probabilistic set membership estimation problems.
\end{abstract}
\end{frontmatter}

%===============================================================================
\section{Introduction}
\label{sec:intro}

In real-world applications, the complexity of the phenomena encountered and the random nature of data makes dealing with uncertainty essential. In many cases, uncertainty arises in the modeling phase, in some others it is intrinsic to both the system and the operative environment, as for instance wind speed and turbulence in aircraft or wind turbine control \cite{Prekopa2013}. 
Hence, it is crucial to include underlying stochastic characteristic of the framework and eventually accept a violation of constraints with a certain probability level, in order to improve the coherence of the model and reality. 
Deriving results in the presence of uncertainty is of major relevance in different areas, including, but not limited to, optimization \cite{sahinidis2004optimization} and robustness analysis \cite{BenNem:98}. However, with respect to robust approaches, where the goal is to determine a feasible solution which is optimal in some sense for \textit{all} possible uncertainty instances , the goal in the stochastic framework is to find a solution that is feasible for \textit{almost} all possible uncertainty realizations, \cite{Calafiore11,Tempo2012_RandAlgForAnalysisAndDesign}.
In several applications, including engineering and finance, where uncertainties in price, demand, supply, currency exchange rate, recycle and feed rate, and demographic condition are common, it is acceptable, up to a certain safe level, to relax the inherent conservativeness of robust constraints enforcing probabilistic constraints. More recently, the method has been used also in unmanned autonomous vehicle navigation \cite{mammarella2018sample,li2019generic} as well as optimal power flow \cite{DabbeneOPF-survey,DabbeneOPF-TCNS}.

In the optimization framework, constraints involving stochastic parameters that are required to be satisfied with a pre-specified probability threshold are called \textit{chance constraints} (CC).
 In general, dealing with CC implies facing two serious challenges, that of stochasticity and of nonconvexity \cite{CCO-survey}. Consequently, while being attractive from a modeling viewpoint, problems involving CC are often computationally intractable, generally shown to be NP-hard, which seriously limits their applicability. However, being able to efficiently solve CC problems remains an important challenge, especially in systems and control, where CC often arise, as e.g. in stochastic model predictive control (SMPC) \cite{matthias1,Mammarella:18:TCST}. 
The scientific community has devoted large research in devising computationally efficient approaches to deal with chance-constraints. We review such techniques in Section~\ref{sec:CSS_overview}, where we highlight  three mainstream approaches:  i) \textit{exact techniques}; ii) \textit{robust approximations}  and iii) \textit{sample-based approximations} .
In this paper, we present what we consider an important step forward in the sample-based approach. We propose a simple and efficient strategy to obtain a probabilistically guaranteed inner approximation of a chance constrained set, with given confidence.

In particular, we describe a two step procedure the involves: i) the preliminary approximation of the chance constraint set by means of a so-called Simple Approximating Set (SAS), ii) a sample-used scaling procedure that allows to properly scale the SAS so to guarantee the desired probabilistic properties. 
The proper selection of a low-complexity SAS allows the designer to easily tune the complexity of the approximating set, significantly reducing the sample complexity. We propose several candidate SAS shapes, grouped in two classes: i) sampled-polytopes; and ii) norm-based SAS. 

The probabilistic scaling approach was presented in the conference papers \cite{alamo2019safe,CCTA2020}.
The present work extends these in several directions: first, we performe here a thorough mathematical analysis the results, providing of all results. Second, the use of norm-based SAS is extended to comprise more general sets (as e.g. , and 
More importantly, we consider here \textit{joint chance constraints}. This choice is motivated by the fact that enforcing joint chance constraints, which have to be satisfied simultaneously, adheres better to some applications, despite the inherent complexity.  
Finally, we present here a second application, besides SMPC, related to probabilistic set-membership identification.

The paper is structured as follows. Section \ref{sec:formulation} provides a general preamble of the problem formulation and of chance constrained optimization, including two motivating examples. An extensive overview on methods for approximating chance constrained sets is reported in Section \ref{sec:CSS_overview} whereas the probabilistic scaling approach has been detailed in Section \ref{sec:scaling}. Section \ref{SAS-sampled-poly} and
Section \ref{sec:normSAS} are dedicated to the definition of selected candidate SAS, i.e. sampled-polytope and norm-based SAS, respectively. Last, in Section \ref{sec:num_ex}, we validate the proposed approach with a numerical example applying our method to a probabilistic set membership estimation problem. Main conclusions and future research directions are addressed in Section \ref{sec:binary}.   

{\small
\subsection{Notation}
Given an integer $N$, $\NATS{N}$ denotes the integers from 1 to~$N$.
Given $z\in \R^s$ and $p \in[1,\infty)$, we denote
by $\|z\|_p$
the $\ell_p$-norm of $z$, and  by
 $\mathbb{B}^s_{p} \doteq \set{z\in \R^s}{\|z\|_p\leq 1}$
 $\ell_p$-norm ball of radius one.
Given integers $k,N$, and parameter $p\in(0,1)$,
the Binomial cumulative distribution function is denoted as
\begin{equation}
\label{eq:bin} 
\Bin(k;N,p) \doteq \Sum{i=0}{k}\conv{N}{i}p^i(1-p)^{N-i}.
\end{equation}
The following notation is borrowed from the field of order statistics \cite{Ahsanullah:13}.
Given a set of $N$ scalars $\gamma_i\in\R^N$, $i\in\NATS{N}$,
we denote $\gamma_{1:N}$ the smallest one, $\gamma_{2:N}$ the second smallest one, and so on and so forth until $\gamma_{N:N}$, which is equal to the largest one.
In this way, given $r\geq 0$ we have that  $\gamma_{r+1:N}$ satisfies that no more than $r$ elements of $\{ \gamma_1, \gamma_2, \ldots, \gamma_N\}$ are strictly smaller than $\gamma_{r+1:N}$.\\
The Chebyshev center of a given set $\X$, denoted as $\mathsf{Cheb}(\X)$, is defined as the center of the largest ball inscribed in $\X$, i.e.\
$$\mathsf{Cheb}(\X)\doteq\arg\min_{\theta_c } \max_{\theta\in\X}\left\{\|\theta-\theta_c \|^2\right\}.$$\\
Given an $\ell_p$-norm  $\|\cdot\|_p$, its dual norm  $\|\cdot\|_{p^*}$
is defined~as 
$$\|c\|_{p^*} \doteq \sup\limits_{z\in\mathbb{B}^s_p} c^\top z, \; \forall c\in \R^s.$$
In particular, the couples $(p,p^*)$: $(2,2)$, $(1,\infty)$, $(\infty,1)$ give raise to dual norms.
}

\section{Problem formulation}
\label{sec:formulation}

Consider a robustness problem, in which the controller parameters and auxiliary variables are parametrized by means of a decision variable vector $\theta$, which is usually referred to as \textit{design parameter} and is restricted to a set $\Theta\subseteq\R^{n_\theta}$. 
Furthermore, the uncertainty vector $\q\in\mathbb{R}^{n_\q}$ represents one of the admissible uncertainty realizations of a random vector with given probability distribution $\Pq$ and (possibly unbounded) support $\Q$. 

This paper deals with the special case where the design specifications can be decoded as a set of  $n_\ell$ uncertain linear inequalities
\begin{equation}\label{eq:ineq}
F(\q)\theta\le  g(\q),
\end{equation}
where
\[
F(\q) = \begin{bmatrix}
       f_1^\top(\q)\\
       \vdots\\
        f_{n_\ell}^\top(\q)      
        \end{bmatrix}
\in\mathbb{R}^{n_\ell\times{n_\theta}},
\quad
g(\q) = \begin{bmatrix}
       g_1(\q)\\
       \vdots\\
        g_{n_\ell}(\q)      
        \end{bmatrix}
        \in\mathbb{R}^{n_\ell},
 \]
 are measurable functions of the uncertainty vector $\q\in\mathbb{R}^{n_\q}$. The inequality in \eqref{eq:ineq} is to be interpreted component-wise, i.e. $$f_\ell(\q) \theta \le g_\ell(\q), \forall \ell\in\NATS{n_\ell}.$$
Furthermore, we notice that each value of $\q$ gives raise to a corresponding set
\begin{equation}\label{Xq}
\X(\q) = \set{\theta\in \Theta}{F(\q)\theta \leq g(\q)}.
\end{equation}
Due to the random nature of the uncertainty vector $\q$, each realization of $\q$ corresponds to a different set of linear inequalities. Consequently, each value of $\q$ gives raise to a corresponding set
\begin{equation}\label{Xq2}
\X(\q) = \set{\theta\in \Theta}{F(\q)\theta \leq g(\q)}.
\end{equation}
In every application, one usually accepts a risk of violating the constraints. While this is often done by choosing the set~$\Q$ appropriately, we can find a less conservative solution by choosing the set $\Q$ to encompass all possible values and characterizing the region of the design space $\Theta$ in which the fraction of elements of $\Q$, that violate the constraints, is below a specified level.
This concept is rigorously formalized by means of the notion of \emph{probability of violation}.

\begin{definition}[Probability of violation] \label{def:prob:violation}
Consider a probability measure ${\rm Pr}_{\Q}$ over $\Q$ and
let $\theta\in \Theta$ be given. The probability of violation of $\theta$ relative to inequality \eqref{eq:ineq} is defined as
$$\viol(\theta)\doteq\Pr_{\Q}\,\{\,F(\q)\theta \not \le  g(\q)\,\}.$$
\end{definition}
\vskip -5mm
Given a constraint on the probability of violation, i.e. \mbox{$\viol(\theta)\leq\varepsilon$}, we denote as (joint) \emph{chance constrained set} of probability $\varepsilon$ (shortly, $\varepsilon$-CCS) the region of the design space for which this probabilistic constraint is satisfied. This is formally stated in the next definition.

\begin{definition}[$\varepsilon$-CCS]
Given $\varepsilon\in(0,1)$, we define the chance constrained set of probability $\varepsilon$ as follows
\begin{equation}
    \label{Xe}
\Xe = \set{\theta\in \Theta}{\viol(\theta)\leq \varepsilon}.
\end{equation}
\end{definition}
\vskip -5mm
Note that the $\varepsilon$-CCS represents the region of the design space~$\Theta$ for which this probabilistic constraint is satisfied and it is equivalently defined as
\begin{equation}
\label{Xe_joint}
\Xe \doteq
\Bigl\{\theta\in\Theta\;:\;
\Pq\left\{
F(\q)\theta \le  g(\q)
\right\}\geq 1-\varepsilon
\Bigr\}.
\end{equation}

\begin{remark}[Joint vs.\ individual CCs]\label{rem-joint}
The constraint $\theta\in\Xe$, with $\Xe$ defined in \eqref{Xe_joint}, describes a \textit{joint chance constraint}. That is, it requires that the joint probability of satisfying the inequality constraint $$F(\q)\theta \le  g(\q)$$ is guaranteed to be greater than the probabilistic level $1-\varepsilon$. We remark that this constraint is notably harder to impose than \textit{individual} CCs, i.e.\ constraints of the form
\begin{eqnarray*}
 \theta\in\X_{\varepsilon_\ell}^\ell\,\, 
 &\!\!\!\!
 \doteq
\!\!\! &
\Bigl\{\theta\in\Theta\,:\,
\Pq\left\{
f_\ell(\q)^\top\theta \le g_\ell(\q)
\right\}\geq 1-\varepsilon_\ell
\Bigr\}, \\
&&\qquad\ell\in\NATS{n_\ell},
\end{eqnarray*}
with $\varepsilon_\ell\in(0,1)$. A discussion on the differences and implications of joint and individual chance constraints may be found in several papers, see for instance \cite{CCO-survey,Miller1965} and references therein.
\end{remark}
\medskip

\begin{example}
A simple illustrating example of the set $\varepsilon$-CCS is shown in Figure \ref{fig:motivatingSphereExample}. The dotted circle is the region of the design space that satisfies all the constraints (the so called robust region), which are tangent to the dotted circle at points uniformly generated. The outer red circle represents the chance constrained set $\Xe$ for the specific value $\varepsilon=0.15$. That is, the red circle is obtained in such a way that every point in it has a probability of violating a random constraint no larger than $0.15$. Note that in this very simple case, the set $\Xe$ can be computed analytically, and turns out to be  a scaled version of the robust set. We observe that the \CCS~is significantly larger than the robust set.

\begin{figure}
\centering
\begin{comment}
\newcommand{\lineAtCirc}[4]{\begin{scope}[circRad=#1,lineAngle=#2,randLLength=#3]
\draw[#4] (#2:#1) -- ++(#2+90:#3);
\draw[#4] (#2:#1) -- ++(#2-90:#3);
\end{scope}
}
\def\circRad{0.1\textwidth}
\begin{tikzpicture}
\def\figWidth{1.3*\circRad}
\def\epsilon{0.15}
\def\outerCircRad{(\circRad/cos(\epsilon*180))}
% coord axis
\draw[-stealth,gray] (-\figWidth,0) -- (1.20*\figWidth,0) node[anchor=north] {$\theta_1$};
\draw[-stealth,gray] (0,-\figWidth) -- (0,1.20*\figWidth) node[anchor=east] {$\theta_2$};
% inner circle (robust set)
\draw[dotted] (0,0) circle (\circRad);
% outer circle (chance constrained set)
\draw[red] (0,0) circle (\outerCircRad);

\def\randHLLength{\circRad}
\pgfmathsetseed{1} % change seed to get different realizations
\foreach \randAngle in {1,2,...,20} % choose number as desired
{\pgfmathsetmacro{\randAngle}{rnd*360}
\draw (\randAngle:\circRad) -- ++(\randAngle+90:\randHLLength);
\draw (\randAngle:\circRad) -- ++(\randAngle-90:\randHLLength);
\clip (-\figWidth,-\figWidth) rectangle (\figWidth,\figWidth);
\lineAtCirc{\circRad}{\randAngle}{\randHLLength}{blue}}
\end{tikzpicture}
\end{comment}
\includegraphics[width=.7\columnwidth]{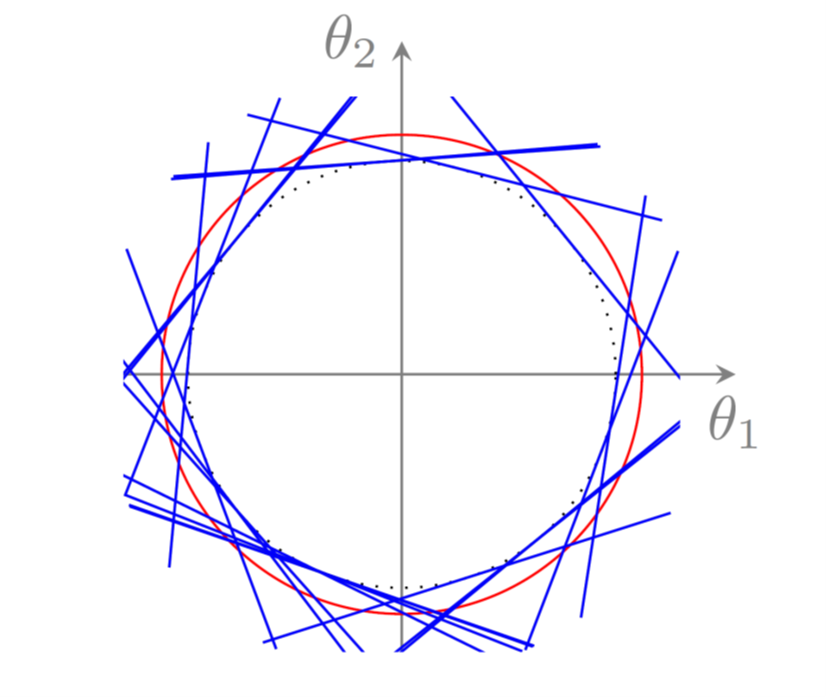}%
\caption{Red circle = $\Xe$, dotted circle = unit circle, blue lines = constraint samples.}
\label{fig:motivatingSphereExample}
\end{figure}
\end{example}
\medskip
\vskip -3mm
Hence, while there exist simple examples for which a closed-form computation of $\Xe$ is possible, as the one re-proposed here and first used in \cite{alamo2019safe}, we remark that this is not the case in general. Indeed, as pointed out in \cite{CCO-survey}, typically the computation of the \CCS~is  extremely difficult, since the evaluation of the probability $\viol(\theta)$ amounts to the computation of a multivariate integral, which is NP-Hard \cite{Khachiyan1989}.

Moreover, the set \CCS~is often nonconvex, except for very special cases. For example, \cite{Prekopa2013,Shapiro-book} show that the solution set of separable chance constraints can be written as the union of cones, which is nonconvex in general. 

\begin{figure}[!h]
\begin{center}
\includegraphics[width=1\columnwidth]{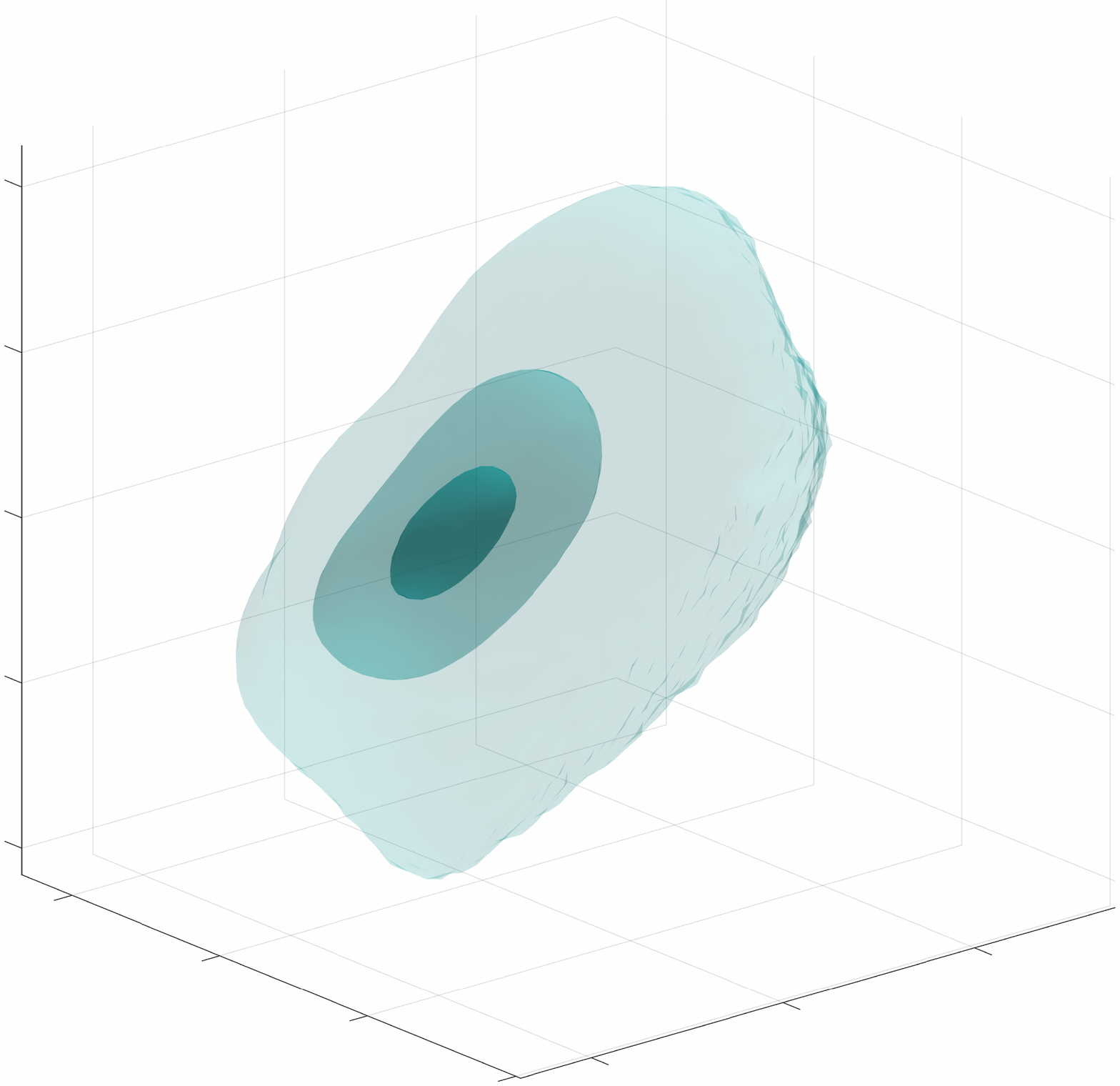}%
\label{f:Xe3D}
\caption{The \CCS~set for $\varepsilon= 0.15$ (smaller set), $\varepsilon= 0.30$ (intermediate set), and $\varepsilon= 0.45$ (larger set). We  observe that all sets are nonconvex, but the nonconvexity is more evident for larger values of $\varepsilon$, corresponding to larger levels of accepted violation, while the set $\Xe$ appears ``almost convex" for small values of $\varepsilon$. This kind of behaviour is in accordance with a recent result that prove convexity of the $\varepsilon$-CCS for values of $\varepsilon$ going to zero, and it is usually referred to as \textit{eventual convexity}~\cite{VanAckooij2015}.
}
\end{center}
\end{figure}

\begin{example}[Example of nonconvex \CCS]
\label{ex:ill}
To illustrate these inherent difficulties, we consider the following three-dimensional example ($n_\theta=3$) with $\q= \left\{\q_1, \q_2\right\}$, where the first uncertainty $\q_1\in\R^3$ is a three-dimensional normal-distributed random vector with zero mean and covariance matrix
\vskip -3mm
%{\small
\[
\Sigma=\left[\begin{array}{ccc}
4.5 &   2.26   & 1.4\\
    2.26 &  3.58  &  1.94\\
    1.4  &  1.94  &  2.19
 \end{array}
\right],
\]
%}%
and the second uncertainty $\q_2\in\R^3$ is a three-dimensional random vector whose elements are uniformly distributed in the interval $[0,1]$. The set of viable design parameters is given by $n_\ell=4$ uncertain linear inequalities of the form
\begin{equation}
F(\q) \theta \le \mathbf{1}_{4},
\quad
 F(\q)=\left[
\begin{array}{cccc}
\q_1 &
\q_2 &
(2\q_1-\q_2) &
\q_1^2
\end{array}
\right]^\top.
\label{Fg_ex}
\end{equation}
The square power $\q_1^2$ is to be interpreted element-wise.

In this case, to obtain a graphical representation of the set $\Xe$, we resorted to gridding the set $\Theta$ and, for each point $\theta$ in the grid, to approximate the probability through a Monte Carlo computation. This procedure is clearly unaffordable for higher dimensions frameworks.
In Figure~\ref{f:Xe3D} we report the plot of the computed \CCS~set for different values of~$\varepsilon$. We observe that the set is indeed nonconvex. 
\end{example}

\subsection{Chance constrained optimization}
Finding an optimal $\theta\in \Xe$ for a given cost function $J:~\mathbb{R}^{n_\theta} \rightarrow \mathbb{R}$, leads to the \textit{chance constrained optimization} (CCO) problem
\begin{equation}\label{CCO}
\min_{\theta \in \Xe} J(\theta),
\end{equation}
where the cost-function $J(\theta)$ is usually assumed to be a convex, often even a quadratic or linear function.

We remark  that the solution of the CCO problem \eqref{CCO} is in general NP-hard, for the same reasons reported before. We also note that several  stochastic optimization problems arising in different application contexts  can be formulated as a CCO. Typical examples are for instance the reservoir
system design problem proposed in \cite{Prekopa1978}, where the problem is to minimize the total building and penalty costs while satisfying demands for all sites and all periods with a given probability, or the cash
matching problem \cite{Dentcheva2004}, where one aims at maximizing the portfolio value at the end of the
planning horizon while covering all scheduled payments with a prescribed probability. 
CCO problems also frequently arise in short-term planning problems in power systems. These optimal power flow (OPF) problems are routinely solved as part of the real-time operation of the power grid. The aim is determining minimum-cost production levels of controllable generators subject to reliably delivering electricity to customers across a large geographical area, see e.g. \cite{DabbeneOPF-survey} and references therein.

In the next subsections, we report two control-related problems which served as motivation of our study.

\subsection{First motivating example: Stochastic MPC}
To motivate the proposed approach, we consider the Stochastic MPC framework proposed in \cite{Mammarella:18:TCST,matthias1}. 
We are given a discrete-time system %
\begin{equation}
x_{k+1} = A(\w_{k})x_{k}+B(\w_{k})u_{k}+a_{\w}(\w_{k}),
\label{eq:sys}
\end{equation}

subject to generic uncertainty $\w_{k} \in \mathbb{R}^{n_{\w}}$,
with state $x_{k} \in \mathbb{R}^{n_x}$, control input $u_{k} \in \mathbb{R}^{n_u}$, and the vector valued function $a_\w(\w_{k})$ representing additive disturbance affecting the system state. The system matrices $A(\w_{k})$ and $B(\w_{k})$, of appropriate dimensions, are (possibly nonlinear) functions of the uncertainty $\w_{k}$ at step $k$. 
For $k=1,2,\ldots$, the disturbances $\w_{k}$ are modeled as realizations of a stochastic process. In particular, $\w_k$ are assumed to be independent and identically distributed (iid) realizations of zero-mean random variables with support $\W\subseteq \mathbb{R}^{n_{\w}}$.
%
%Moreover, we assume that  let $\mathbb{G}=\left\{(A(\w_{k}),B(\w_{k}),a_{w}(\w_{k})\right\}_{\w_{k}\in \mathbb{W}}$, a polytopic outer approximation with $N_c$ vertexes $\bar{\mathbb{G}}\doteq co\left\{A^{j},B^{j},a_{w}^{j}\right\}_{j\in \mathbb{N}_{1}^{N_{c}}}\supseteq \mathbb{G}$ exists and is known.
%
%We cannotice that the system can be augmented by a filter to model a specific stochastic processes of interest.  
%
Note that the presence of both additive and multiplicative uncertainty, combined with the nonlinear dependence on the uncertainty, renders the problem particularly arduous. Furthermore, we remark that the system representation in \eqref{eq:sys} is very general, and encompasses, among others, those in \cite{matthias1,Mammarella:18:TCST,matthias2}. 

Given the model~\eqref{eq:sys} and a realization of the state $x_k$ at time $k$, state predictions $\t$ steps ahead are random variables as well and are denoted $x_{\t|k}$, to differentiate it from the realization $x_{\t+k}$. Similarly $u_{\t|k}$ denotes predicted inputs that are computed based on the realization of the state $x_k$.

Contrary to \cite{matthias1,Mammarella:18:TCST,matthias2}, 
where the system dynamics were subject to individual state and input chance constraints, here we take a more challenging route, and 
 we consider \textit{joint state} and \textit{input chance constraints} of the form
\footnote{The case where one wants to impose \textit{hard} input constraints can be also be formulated in a similar framework, see e.g.~\cite{matthias1}.}
\begin{equation}
\Pwb \left\{  H_x x_{\t|k} + H_u  u_{\t|k} \le  \mathbf{1}_{n_\t}   |x_k \right\}
\ge 1-\varepsilon, 
\label{SMPC-CC}
\end{equation}
with $\t\in  \{0, \ldots, T-1\}$, $\varepsilon \in (0,1)$, and $H_x\in \R^{n_\ell \times n_x}$, $H_u \in \R^{n_\ell \times n_u}$.

The probability $\Pwb$ is measured with respect to the sequence ${\pmb{\sigma}}=\{\w_\t\}_{\t>k}$. Hence, equation~\eqref{SMPC-CC} states that the probability of violating the linear constraint 
$H_x x+H_u u \le 1$ for any future realization of the disturbance should not be larger than $\varepsilon$.

The objective is to derive an asymptotically stabilizing control law for the system~\eqref{eq:sys} such that, in closed loop, the constraint~\eqref{SMPC-CC} is satisfied.
Following the approach in \cite{Mammarella:18:TCST}, a stochastic MPC algorithm is considered to solve the constrained control problem. The approach is based on repeatedly solving a stochastic optimal control problem over a finite, moving horizon, but implementing only the first control action. 
The design parameter $\theta$ is then given by the control sequence $\mathbf{u}_{k} = (u_{0|k}, u_{1|k}, ..., u_{T-1|k})$ and the prototype optimal control problem to be solved at each sampling time $k$ is defined by the cost function
\begin{eqnarray}    \label{SMPC-cost}
 \lefteqn{J_T(x_k,\mathbf{u}_{k}) = }\\
 & \mathbb{E}\left\{ \sum_{t=0}^{T-1} \left( x_{\t|k}^\top Q x_{\t|k} + u_{\t|k}^\top Ru_{t|k} \right) + x_{T|k}^\top P x_{T|k} ~|~ x_k\right\},\nonumber
\end{eqnarray}
with $Q\in\mathbb{R}^{n_x \times n_x}$, $Q\succeq 0$, $R\in\mathbb{R}^{n_u\times n_u}$, $R \succ 0$, and appropriately chosen $P \succ 0$, subject to the system dynamics~\eqref{eq:sys} and constraints~\eqref{SMPC-CC}.

The online solution of the stochastic MPC problem remains a challenging task but several special cases, which can be evaluated exactly, as well as methods to approximate the general solution have been proposed in the literature. The approach followed in this work was first proposed in~\cite{matthias1,Mammarella:18:TCST}, where an offline sampling scheme was introduced. Therein, with a prestabilizing input parameterization
\begin{equation}
u_{\t|k}=Kx_{\t|k}+v_{\t|k},
\label{eq:prestabilizingInput}
\end{equation}
with suitably chosen control gain $K\in\mathbb{R}^{n_u\times n_x}$ and new design parameters $v_{\t|k} \in \R^{n_u}$, equation~\eqref{eq:sys} is solved explicitly for the predicted states $x_{1|k},\ldots,x_{T|k}$ and predicted inputs $u_{0|k},\ldots,u_{T-1|k}$.
In this case, the expected value of the finite-horizon cost~\eqref{SMPC-cost} can be evaluated \textit{offline}, leading to a quadratic cost function of the form
\begin{equation}
J_{T}(x_{k},
\mathbf{v}_{k})=
\begin{bmatrix}
        x_{k}^\top & \textbf{v}_{k}^\top  &        \textbf{1}_{n_x}^\top
        \end{bmatrix}\tilde{S}
\begin{bmatrix}
        x_{k} \\
        \textbf{v}_{k} \\
        \textbf{1}_{n_x}\\
        \end{bmatrix}
        \label{SMPC-cost_new}
\end{equation}
in the deterministic variables $\mathbf{v}_{k} = (v_{0|k}, v_{1|k}, ..., v_{T-1|k})$ and $x_{k}$.

Focusing now on the constraint definition, we notice that by introducing the uncertainty sequence $\pmb{\sigma}_k=\{\w _t\}_{t=k,...,k+T-1}$, we can rewrite the joint chance constraint defined by equation~\eqref{SMPC-CC} as 
\begin{align}
&    \Xe^{\textsc{smpc}}= 
    \left\{    \    \begin{bmatrix}
    x_k\\
    \mathbf{v}_k
     \end{bmatrix} \in\mathbb{R}^{n_x+n_uT} ~:~\right.\nonumber\\ 
&\Bigl.
  \mathsf{Pr}_{\pmb{\w}_k}
   \left\{   
    \begin{bmatrix}
     f_\ell^x (\pmb{\w}_k)\\
     f_\ell^v(\pmb{\w}_k)
    \end{bmatrix}^\top
    \begin{bmatrix}
    x_k\\
    \mathbf{v}_k
     \end{bmatrix}
     \leq 1, \ell\in\NATS{n_\ell} \right\} \geq 1-\varepsilon \Bigr\},
    \label{SMPC-Xe}
\end{align}
with $f_\ell^x:\R^{n_\w}\to\R^{n_x},f_\ell^v:\R^{n_\w}\to\R^{n_uT}$ being known functions of the  sequence of random variables $\pmb{\w}_k$. We remark that, in the context of this paper, neither the detailed derivation of the cost matrix $\tilde{S}$ in \eqref{SMPC-cost_new}  nor that of $f_\ell^v, f_\ell^x$ are relevant for the reader, who can refer to \cite[Appendix A]{Mammarella:18:TCST} for details. Note that, by defining 
$\theta=[x_k^\top,\mathbf{v}_k^\top]^\top$, \eqref{SMPC-Xe} is given in the form of \eqref{Xe} .

As discussed in \cite{matthias1}, obtaining a good and simple enough approximation of the set $\Xe^{\textsc{smpc}}$ is extremely important for online implementation of SMPC schemes. In particular, if we are able to replace the set $\Xe^{\textsc{smpc}}$ by a suitable inner approximation, we would be able to guarantee probabilistic constraint satisfaction of the ensuing SMPC scheme. On the other hand, we would like this inner approximation to be simple enough, so to render the online computations fast enough.

\subsection{Second motivating example: probabilistic set membership estimation}
\label{sub:sec:Second:Motivating:Set:membeship}
Suppose that there exists $\bar{\theta}\in \Theta$ such that   
$$ |y-\bar{\theta}^T\varphi(x)|\leq \rho,\; \forall (x,y)\in \Q \subseteq \R^{n_x}\times \R,$$
where $\varphi:\R^{n_x}\to \R^{n_\theta}$ is a (possibly non-linear) regressor function, and $\rho>0$ accounts for modelling errors. 
The (deterministic) set membership estimation problem, see \cite{vicino1996sequential}, \cite{bravo2006bounded}, consists of computing the set of parameters $\theta$ that satisfy 
the constraint $$|y-\theta^T\varphi(x)|\leq \rho$$ for all possible values of $(x,y)\in \Q$.
In the literature, this set is usually referred to as the \textit{feasible parameter set}, that is
\begin{equation}\label{equ:FPS} 
\FPS \doteq
\set{\theta\in \Theta}{
|y-\theta^T\varphi(x)|\leq \rho, \;\forall (x,y)\in \Q}.
\end{equation}
If, for given $w=(x,y)$, we define the set
\[
\X(\q) = \set{\theta\in \Theta}{|y-\theta^T \varphi(x) |\leq \rho},
\]
then the feasible parameter set $\FPS$  can be rewritten as
\[ 
\FPS = \set{\theta\in \Theta}{\theta \in \X(\q), \; \forall w\in \Q}.
\]
The deterministic set membership problem suffers from the following limitations in real applications:
i) due to the possible non-linearity of $\varphi(\cdot)$, checking if a given $\theta\in \Theta$ satisfies the constraint $\theta\in \X(\q)$, for every $w\in \Q$, is often a difficult problem; ii)
in many situations, only samples of $\Q$ are available: thus, the robust constraint cannot be checked and only outer bounds of $\FPS$ can be computed; and iii) because of outliers and possible non finite support of $\Q$, set $\FPS$ is often empty (especially for small values of $\rho$).

If a probability distribution is defined on $\Q$, the probabilistic set membership estimation problem is that of characterizing the set of parameters $\theta$ that satisfy
\[
 \Pq \{ | y-\theta^T \varphi(x)| \leq \rho \} \geq 1-\epsilon,
 \]
for a given probability parameter $\epsilon\in(0,1)$.
Hence, we can define $\FPS_\epsilon$ the set of parameters that satisfy the previous probabilistic constraint, that is,
\[
 \FPS_\epsilon =\set{\theta\in \Theta}{\Pq\{\theta\in \X(\q)\} \geq 1-\epsilon}.
 \]
It is immediate to notice that this problem fits in the formulation proposed in this section: It suffices to define 
$$F(\q) = \left[\begin{array}{c} \varphi^T(x) \\ -\varphi^T(x)\end{array}\right], \; 
g(\q) = \left[\begin{array}{c} \rho+y \\ \rho-y\end{array}\right].$$

\subsection{Chance constrained approximations}
Motivated by the discussion above, we are ready to formulate the main problem studied in this paper.
\medskip

\begin{problem}[\CCS~approximation]\label{problem_CCS}
Given the set of linear inequalities \eqref{eq:ineq}, and a violation parameter $\varepsilon$, find an inner approximation of the set $\Xe$. The approximation should be: i) simple enough, ii) easily computable.
\end{problem}

A solution to this problem is provided in the paper. In particular, regarding i), we present a solution in which the approximating set is represented by \textit{few linear inequalities}. Regarding ii), we propose a computationally efficient procedure for its construction (see Algorithm~1).

Before presenting our approach, in the next section we provide a brief literature overview of different methods presented in the literature to construct approximations of the \CCS~set.
\medskip
\vskip -1cm
\section{Overview on different approaches to $\varepsilon$-CCS approximations}
\label{sec:CSS_overview}
The construction of computational efficient approximations to $\varepsilon$-CCS is a long-standing problem. In particular, the reader is referred to the recent work \cite{CCO-survey}, which provides a rather complete discussion on the topic, and covers the most recent results. The authors distinguish three different approaches, which we very briefly revisit here.

\subsection{Exact techniques}
In some very special cases, the $\varepsilon$-CCS is convex and hence the CCO problem admits a unique solution. This is the case, for instance, of \textit{individual} chance constraints with~$\q$  being Gaussian \cite{Kataoka1963}.
Other important examples of convexity of the set $\Xe$ involve log-concave distribution \cite{Prekopa2013,Prekopa1971}. General sufficient conditions on the convexity of chance constraints may be found in \cite{Lagoa1999,CalafioreEl2006,Henrion2008,VanAckooij2015}. However, all these cases are very specific and hardly extend to joint chance constraints considered on this work. 

\subsection{Robust techniques}
A second class of approaches consist in finding \textit{deterministic} conditions that allow to construct a set $\underline{\X}$, which is a guaranteed inner convex approximation of the probabilistic set $\Xe$. The classical solution consists in the applications of Chebyshev-like inequalities, see e.g. \cite{hewing2018stochastic,yan2018stochastic}. More recent techniques, which are proved particularly promising, involve robust optimization \cite{BenNem:98}, as the convex approximations introduced in \cite{Nemirovski06}. A particular interesting convex relaxation involves the so-called Conditional Value at Risk (CVaR), see \cite{ChenCVAR} and references therein. Finally, we point out some recent techniques based on  polynomial moments relaxations \cite{LagoaCCO,LasserreCCO}. Nonetheless, it should be remarked that these techniques usually suffer from conservatism and computational complexity issues, especially in the case of joint chance constraints.

\subsection{Sample-based techniques}
\label{sec:rand}
In recent years, a novel approach to approximate chance constraints, based on random sampling of the uncertain parameters, has gained popularity, see e.g.~\cite{Calafiore11,Tempo2012_RandAlgForAnalysisAndDesign} and references therein. Sampling-based techniques are characterized by the use of a finite number $N$ of iid samples of the uncertainty
$\left\{ \q^{(1)}, \q^{(2)}, \ldots, \q^{(N)}\right\}$
drawn according to a probability distribution $\Pq$.
To each sample $\q^{(i)}, i\in\NATS{N}$, we can associate the following \textit{sampled set} 
\begin{equation}
\label{eq:Xi}
\X(\q^{(i)})=\set{\theta\in\Theta}{F(\q^{(i)})\theta\le g(\q^{(i)})},
\end{equation}
sometimes referred to as \textit{scenario}, since it represents an observed instance of our probabilistic constraint. 

Then, the scenario approach considers the CCO problem~\eqref{CCO} and approximates its solution through the following \textit{scenario problem}
\begin{align}\label{eq:scen}
&\theta^*_{sc}=\arg\min J(\theta)\\
&\text{subject to } \theta\in\X(\q^{(i)}), i\in\NATS{N}. \nonumber
\end{align}
We note that, if the function $J(\theta)$ is convex, problem~\eqref{eq:scen} becomes a linearly constrained convex program, for which very efficient solution approaches exist. A fundamental result \cite{Calafiore06,Campi08,CalafioreSIAM10,Campi11} provides a probabilistic certification of the constraint satisfaction for the solution to the scenario problem. In particular, it is shown that, under some mild assumptions (non-degenerate problem), we have
\begin{equation}
\label{eq:viol_scen}
\PqN\left\{\viol(\theta^*_{sc})>\varepsilon\right\} \le \Bin(n_\theta-1;N,\varepsilon),
\end{equation}
where the probability in \eqref{eq:viol_scen} is measured with respect to the samples $\{\q^{(1)}, \q^{(2)}, \ldots, \q^{(N)}$\}. Moreover, the bound in \eqref{eq:viol_scen} is shown to be tight. Indeed, for the class of so-called fully-supported problems, the bound holds with equality, i.e. the Binomial distribution $\Bin(n_\theta-1;N,\varepsilon)$ represents the exact probability distribution of the violation probability~\cite{Campi08}.

A few observations are at hand regarding the scenario approach and its relationship with Problem~\ref{problem_CCS}. First, if we define the 
 \textit{sampled constraints set} as
\begin{equation}\label{eq:XN}
    \X_N\doteq\bigcap_{i=1}^{N}\X(\q^{(i)}),
\end{equation}
we see that the scenario approach consists in approximating the constraint $\theta\in\Xe$ in \eqref{CCO} with its sampled version $\theta\in\X_N$. On the other hand, it should be remarked that the scenario approach cannot be used to derive any guarantee on the relationship existing between $\X_N$ and $\Xe$. Indeed, the nice probabilistic property in \eqref{eq:viol_scen} holds \textit{only for the optimum of the scenario program} $\theta^*_{sc}$. This is a fundamental point, since the scenario results build on the so-called support constraints, which are defined for the optimum point $\theta^*_{sc}$ only.

On the contrary, in our case we are interested in establishing a direct relation (in probabilistic terms) between the set $\X_N$ and the \CCS~$\Xe$. This is indeed possible, but needs to resort to results based on Statistical Learning Theory \cite{Vapnik98}, summarized in the following lemma.

\begin{lemma}[Learning Theory bound]
\label{lem:LT}
Given probabilistic levels $\delta\in(0,1)$ and $\varepsilon\in(0,0.14)$, if the number of samples $N$ is chosen so that $N\ge N_{LT}$, with
\begin{equation}
\label{eq:Ntilde}
N_{LT}\doteq \frac{4.1}{\varepsilon}\Big(\ln \frac{21.64}{\delta}+4.39n_\theta\,\log_{2}\Big(\frac{8en_\ell}{\varepsilon}\Big)\Big),
\end{equation}
then 
$    \PqN\left\{\X_N\subseteq \Xe\right\}\geq 1-\delta$.
\end{lemma}

The lemma, whose proof is reported in Appendix~\ref{app:proof-Lemma1}, is a direct consequence of the results on VC-dimension of the so-called $(\alpha,k)$-Boolean Function, given in \cite{alamo2009randomized}.

\begin{remark}[Sample-based SMPC]
\label{rem:SMPC-sampled} The learning theory-based approach discussed in this section has been applied in \cite{matthias1} to derive an \emph{offline} probabilistic inner approximation of the chance constrained set $\Xe^{\textsc{smpc}}$ defined in \eqref{SMPC-Xe}, considering individual chance constraints. In particular, the
bound \eqref{eq:ineq} is a direct extension to the case of joint chance constraints of the result proved in \cite{matthias1}. Note that since we are considering multiple constraints at the same time (like in \eqref{eq:ineq}), the number of constraints $n_\ell$ enters into the sample size bound.
To explain how the SMPC design in \cite{matthias1}  extends to the joint chance constraints framework, we briefly recall it.

First, we extract \textit{offline} (i.e.~when designing the SMPC control) $N$ iid\ samples of the uncertainty, $\pmb{\w}_k^{(i)}$ of $\pmb{\w}_k$, and we consider the sampled set
\begin{align}
   \X^{\textsc{smpc}}(\pmb{\w}_k^{(i)})= 
    \Biggl\{\ \begin{bmatrix}
    x_k\\
    \mathbf{v}_k
     \end{bmatrix} :\nonumber
    \begin{bmatrix}
     f_\ell^x (\pmb{\w}_k^{(i)})\\
     f_\ell^v(\pmb{\w}_k^{(i)})
    \end{bmatrix}^\top
    \begin{bmatrix}
    x_k\\
   \mathbf{v}_k
     \end{bmatrix}
     \leq 1,
\Biggl.
   \ell\in\NATS{n_\ell}\Biggr\},
\end{align}
and $\X_N^{\textsc{smpc}}\doteq\bigcap_{i=1}^{N}\X^{\textsc{smpc}}(\pmb{\w}_k^{(i)})$. Then, applying Lemma~ \ref{lem:LT} with $n_\theta=n_x+n_uT$,
we conclude that if we extract $N\ge N_{LT}^{\textsc{smpc}}$ samples, it is guaranteed that, with probability at least $1-\delta$, the sample approximation $\X_N^{\textsc{smpc}}$ is a subset of the original chance constraint $\Xe^{\textsc{smpc}}$.
Exploiting these results, the SMPC problem can be approximated conservatively by the linearly constrained quadratic program
\begin{align}
&\min_{\mathbf{v}_k} ~J_T(x_k, \mathbf{v}_k) \textrm{ subject to }  (x_k, \mathbf{v}_k) \in \X_N^{\textsc{smpc}}.
\end{align}
Hence the result reduces the original stochastic optimization program to an efficiently solvable quadratic program. This represents an undiscussed advantage, which has been demonstrated for instance in  \cite{Mammarella:18:TCST}. 
On the other hand, it turns out that  the ensuing number of linear constraints,  equal to
$n_\ell\cdot N_{LT}^{\textsc{smpc}}$
may still be too large. For instance, even for a moderately sized MPC problem with $n_x=5$ states, $n_u=2$ inputs, prediction horizon of $T=10$, simple interval constraints on states and inputs (i.e.~$n_\ell=2n_x+2n_u=14$), and
for a reasonable choice of probabilistic parameters, i.e. $\varepsilon=0.05$ and $\delta=10^{-6}$, we get $N_{LT}^{\textsc{smpc}}=114,530$, which in turn corresponds to more than $1.6$ million linear inequalities. For this reason, in~\cite{matthias1} a post-processing step was proposed to remove redundant constraints. While it is indeed true that all the cumbersome computations may be performed offline, it is still the case that, in applications with stringent requirements on the solution time, the final number of inequalities may easily become unbearable.
\end{remark}

Remark \ref{rem:SMPC-sampled} motivates the approach presented in the next section, which builds upon the results presented in~\cite{alamo2019safe}. We show how the probabilistic scaling approach directly leads to approximations of user-chosen complexity, which can be directly used in applications instead of creating the need for a post-processing step to reduce the complexity of the sampled set.

\section{The Probabilistic Scaling Approach}
\label{sec:scaling}
We propose a novel sample-based approach, alternative to the randomized procedures proposed so far, which allows to maintain the nice probabilistic features of these techniques, while at the same time providing the designer with a way of tuning the complexity of the approximation.

The main idea behind this approach consists of first obtaining a simple initial approximation of the \textit{shape} of the probabilistic set $\Xe$ by exploiting \textit{scalable simple approximating sets} (Scalable SAS) of the form
\begin{equation}
  \label{SASgamma} 
  \Su(\gamma) = \theta_c \oplus \gamma \Su.
\end{equation}
These sets are described by a center point $\theta_c $ and a low-complexity shape set $\Su$. The center $\theta_c$ and the shape $\Su$ constitute the \textit{design parameters} of the proposed approach. By appropriately selecting the shape $\Su$, the designer can control the complexity of the approximating set. 

Note that we do not ask this initial set to have \textit{any} guarantee of probabilistic nature. What we ask is that this set is being able to ``capture" somehow the shape of the set~$\Xe$. Recipes on a possible procedure for constructing this initial set are provided in section~\ref{SAS-sampled-poly}. 
The set $\Su$ constitutes the starting point of a scaling procedure, which allows to derive a probabilistic guaranteed approximation of the \CCS, as detailed in the next section.
In particular, we show how an optimal scaling factor $\gamma$ can be derived so that the set~\eqref{SASgamma} is guaranteed to be an inner approximation of $\Xe$ with the desired confidence level $\delta$. We refer to the set $\Sug$ as \textit{Scalable SAS}.

\subsection{Probabilistic Scaling}\label{sec:prob:scaling}

In this section, we address the problem of how to scale the set $\Sug$ around its center $\theta_c $ to guarantee, with confidence level $\delta\in(0,1)$, the inclusion of the scaled set into $\Xe$. Within this sample-based procedure we assume that $N_\gamma$ iid samples $\{\q^{(1)},\ldots, \q^{(N_\gamma)}\}$ are obtained from $\Pr_{\Q}$ and based on these, we show how to obtain a scalar $\bar\gamma>0$ such that
$$ \Pr_{\Q^{N_\gamma}} \{\Su(\bar\gamma) \subseteq \Xe\} \geq 1-\delta.$$
To this end, we first define the scaling factor associated to a given realisation of the uncertainty.

\begin{definition}[Scaling factor]
\label{def-scaling}
  Given a Scalable SAS $\Sug$, with given center $\theta_c $ and shape $\Su \subset \Theta$, and a realization $\q\in\Q$, we define the scaling factor of $\Sug$ relative to $\q$ as
$$
\scale(\q) \doteq \bsis{cc} 0 & \,\,\,\mbox {if} \; \theta_c \not\in \X(\q) \\
 \max\limits_{\Sug \subseteq\X(\q)}  \gamma & \,\,\,\mbox{otherwise}. 
 \esis$$
 with $\X(\q)$ defined as in \eqref{eq:Xi}.
\end{definition}
 \medskip
  
That is $\scale(\q)$ represents the maximal scaling that can be applied to $\Sug=\theta_c \oplus\gamma\Su$ around the center $\theta_c$ so that $\Sug\subseteq\X(\q)$. 
The following theorem states how to obtain, by means of sampling, a scaling factor $\bar\gamma$ that guarantees, with high probability, that $\Su(\bar\gamma)\subseteq\Xe$.
\smallskip

\begin{theorem}[Probabilistic scaling]
\label{th:prop:scaling}
Given a candidate Scalable SAS \,$\Su(\gamma)$, with $\theta_c\in\Xe$, accuracy parameter $\varepsilon\in(0,1)$, confidence level $\delta\in (0,1)$, and a discarding integer parameter $r\geq 0$, let $N_\gamma$ be chosen such that
\begin{equation}
\label{eq:Bin}
\Bin(r;N_\gamma,\varepsilon)\leq \delta.
\end{equation}
Draw $N_\gamma$ iid samples $\{ \q^{(1)}, \q^{(2)}, \ldots, \q^{(N_\gamma)} \}$
from distribution $\Pr_\Q$, 
compute the corresponding scaling factor
\begin{equation}
 \gamma_i \doteq \scale(\q^{(i)}),
\end{equation}
for $i\in\NATS{N_\gamma}$ according to Definition~\ref{def-scaling}, and let $\bar\gamma =  \gamma_{1+r:N_\gamma} $. Then, with probability no smaller than $1-\delta$, 
$$\Su(\bar\gamma)= \theta_c \oplus \bar\gamma \Su \subseteq  \Xe.$$
\end{theorem}
\vskip -3mm
\noindent\textit{Proof:} If $\bar\gamma=0$, then we have $\Su(\bar\gamma)\equiv \theta_c\in\Xe$.
Hence,  consider $\bar\gamma >0$. From Property \ref{prop:scale:sets} in Appendix \ref{app2}, we have that $\bar\gamma0$
satisfies, with probability no smaller than $1-\delta$, that
$\Pr_{\Q} \{\Sug \not \subseteq \X(\q)\} \leq \varepsilon$.
Equivalently,
$ \Pr_{\Q} \{\Sug \subseteq \X(\q)\} > 1-\varepsilon.$
This can be rewritten as
$ \Pr_{\Q} \{F(\q)^\top\theta \le g(\q), \;\; \forall \theta \in \Sug \} > 1-\varepsilon, $
and it implies that the probability of violation in $\theta_c \oplus\bar\gamma \Su$ is no larger than $\varepsilon$, with probability no smaller than $1-\delta$. 
\hfill\qedsymbol

%\FRG{Teo, should we assume that $\Xe$ is nonempty with nonempty interior?}
In the light of the theorem above, from now on we will assume that the Scalable SAS is such that $\theta_c\in\Xe$.
The above result leads to the following
simple algorithm, in which we summarise the main steps for constructing the scaled set, and we provide an explicit way of determining the discarding parameter $r$.

\begin{algorithm}[H]
\caption{Probabilistic SAS Scaling}
\label{alg:scaling}
\begin{algorithmic}[1]
\State 
Given a candidate Scalable SAS $\Su(\gamma)$, and
 probability levels $\varepsilon$ and $\delta$, choose
\begin{equation}
N_\gamma \ge  \frac{7.47}{\varepsilon} \ln\frac{1}{\delta}\quad\text{ and }\quad
r=\left\lfloor \frac{\varepsilon N_\gamma}{2}\right\rfloor.
\label{N_alg}    
\end{equation}
\State Draw $N_\gamma$ samples of the uncertainty 
$\q^{(1)},\ldots,\q^{(N_\gamma)}$
\For  {$i=1$ to $N_\gamma$}
\State Solve the optimization problem
\vskip -7mm
\begin{align}
\label{eq:opt-alg1}
    \gamma_i \doteq &\max_{\Sug \subseteq \X(\q^{(i)})} \gamma 
\end{align}
\EndFor
\State
Return $\bar\gamma=\gamma_{1+r:N_\gamma}$, the $(1+r)$-th smallest value of~$\gamma_i$.
\end{algorithmic}
\label{Algo1}
\end{algorithm}
\vskip -5mm
A few comments are in order regarding the algorithm above. In step 4, for each uncertainty sample $\q^{(i)}$ one has to solve an optimization problem, which amounts to finding the largest value of $\gamma$ such that $\Sug$ is contained in the set $\X(\q^{(i)})$ defined in \eqref{eq:Xi}. If the SAS is chosen accurately, we can show that this problem is convex and computationally very efficient: this is discussed in Section \ref{SAS-sampled-poly}. Then, in step~6, one has to re-order the set 
$\{ \gamma_1, \gamma_2, \ldots, \gamma_{N_\gamma}\}$ so that the first element is the smallest one, the second element is the second smallest one, and so on and so fort, and then return the $r+1$-th element of the reordered sequence. The following Corollary applies to Algorithm~\ref{Algo1}.
\begin{corollary}
\label{lemma:scaling}
Given a candidate SAS set in the form $\Sug= \theta_c \oplus\gamma \Su$, assume that $\theta_c \in\Xe$. Then, Algorithm~\ref{Algo1} guarantees that 
$\Su(\bar\gamma)\subseteq \Xe$
with probability at least $1-\delta$.
\end{corollary}
\noindent\textit{Proof:}
The result is a direct consequence of Theorem \ref{th:prop:scaling}, which guarantees that, for given $r\ge 0$, $\mathsf{Pr}\{\Sug \subseteq \Xe\}$ is guaranteed if the scaling is performed on a number of samples 
satisfying \eqref{eq:Bin}.
From \cite[Corollary 1]{Alamo:15}) it follows that, in order to satisfy \eqref{eq:Bin}
it suffices to take $N_\gamma$ such that
\begin{equation}
\label{eq:Ngamma}
N_\gamma\geq \frac{1}{\varepsilon} \left( r+\ln\frac{1}{\delta}+\sqrt{2r\ln\frac{1}{\delta}}\right).\end{equation} 

Since $r=\lfloor \frac{\varepsilon N}{2} \rfloor$, we have that $r \leq  \frac{\varepsilon N}{2}$. Thus, inequality \eqref{eq:Ngamma} is satisfied if 
\begin{equation*}
\begin{aligned}
N_\gamma &\geq& \frac{1}{\varepsilon} \left( \frac{\varepsilon N_\gamma}{2}+\ln\frac{1}{\delta}+\sqrt{\varepsilon N_\gamma\ln\frac{1}{\delta}}\right)\\
&=&
 \frac{N_\gamma}{2}+\frac{1}{\varepsilon} \ln\frac{1}{\delta}+\sqrt{N_\gamma\frac{1}{\varepsilon} \ln\frac{1}{\delta}}.
\end{aligned}
\end{equation*}
Letting $\nabla\doteq \sqrt{N_\gamma}$ and $\alpha\doteq \sqrt{\frac{1}{\varepsilon} \ln\frac{1}{\delta}}$\footnote{Note that both quantities under square root are positive.}, the above inequality rewrites
$
\nabla^2-2\alpha\nabla -2\alpha^2 \ge 0,$
which has unique positive solution $\nabla\ge (1+\sqrt{3})\alpha$. In turn, this rewrites as 
$$
N_\gamma \ge  \frac{(1+\sqrt{3})^2}{\varepsilon} \ln\frac{1}{\delta}.$$
The formula \eqref{N_alg} follows by observing that $(1+\sqrt{3})^2<~7.47$.
\hfill\qedsymbol

In the next sections, we provide a ``library" of possible candidates SAS shapes. We remind that these sets need to comply to two main requirements: i) being a simple and low-complexity representation; and ii) being able to capture the original shape of the $\varepsilon$-CCS. Moreover, in the light of the discussion after Algorithm \ref{Algo1}, we also ask these sets to be convex. 

\section{Candidate SAS: Sampled-polytope}
\label{SAS-sampled-poly}
First, we note that the most straightforward way to design a candidate SAS is again to recur to a sample-based procedure: we draw a fixed number $N_S$ of ``design" uncertainty samples%
\footnote{These samples are denoted with a tilde to distinguish them from the samples used in the probabilistic scaling procedure.}
$
\{\tilde\q^{(1)},\ldots,\tilde\q^{(N_S)}\}$,
and construct an initial sampled approximation by introducing the following \textit{sampled-polytope SAS}
\begin{equation}
\label{eq:sampledSAS}
    \Su_{N_S}=\bigcap_{j=1}^{N_S}\X(\tilde\q^{(j)}).
\end{equation}

Note that the sampled polytope $\Su_{N_S}$, by construction, is given by the intersection of $n_\ell N_S$ half-spaces. Hence, we observe that this approach provides  very precise control on the final complexity of the approximation, through the choice of the number of samples $N_S$. However, it is also clear that a choice for which $N_S<<N_{LT}$ implies that the probabilistic properties of $\Su_{N_S}$ before scaling will be very bad.
However, we emphasize again that this initial geometry doesn't have nor require any probabilistic guarantees, which are instead provided by the probabilistic scaling discussed in Section \ref{sec:prob:scaling}. It should be also remarked  that this is only one possible heuristic. For instance, along this line one could as well draw many samples and then apply a clustering algorithm to boil it down to a desired number of samples.

We remark that, in order to apply the scaling procedure, we need to define a \textit{center} around which to apply the scaling procedure. To this end, we could compute the so-called Chebyshev center, defined as the center of largest ball inscribed in $\Su_{N_S}$, i.e.\ $\theta_{c}=\mathsf{Cheb}(\Su_{N_S})$. 
We note that computing the Chebyshev center of a given polytope is an easy convex optimization problem, for which efficient algorithms exist, see e.g.~\cite{Boyd04}. A possible alternative would be the analytic center of $\Su_{N_S}$, whose computation is even easier (see \cite{Boyd04} for further details). Once the center $\theta_c $ has been determined, the scaling procedure can be applied to the set $\Su_{N_S}(\gamma)\doteq \theta_c \oplus \gamma \{\Su_{N_S}\ominus \theta_c \}$. Note that the center needs to be inside $\Xe$. Aside for that, the choice of $\theta_c$ only affects the goodness of the shape, but we can never know a priori if the analytic center is a better choice than any random center in $\Xe$.
\begin{figure*}[!ht]
\subfigure[$\Su_{N_S}$ with $N_S=100$. $\rightarrow$ $\gamma=0.8954$]{\includegraphics[height=6cm]{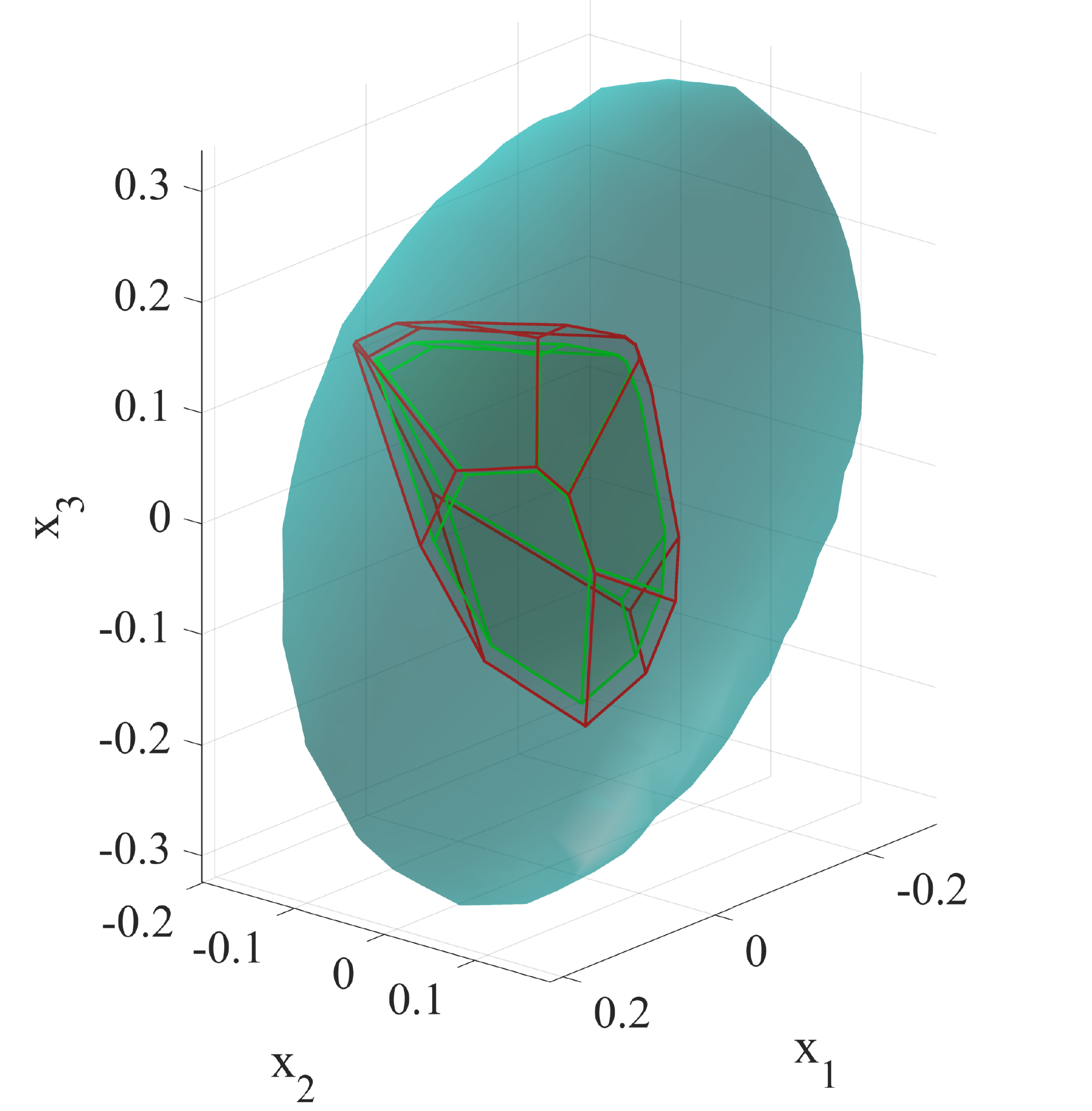}%
\label{f:samplepoly100}}
\subfigure[$\Su_{N_S}$ with $N_S=1,000$. $\rightarrow$ $\gamma=1.2389$]{\includegraphics[height=6cm]{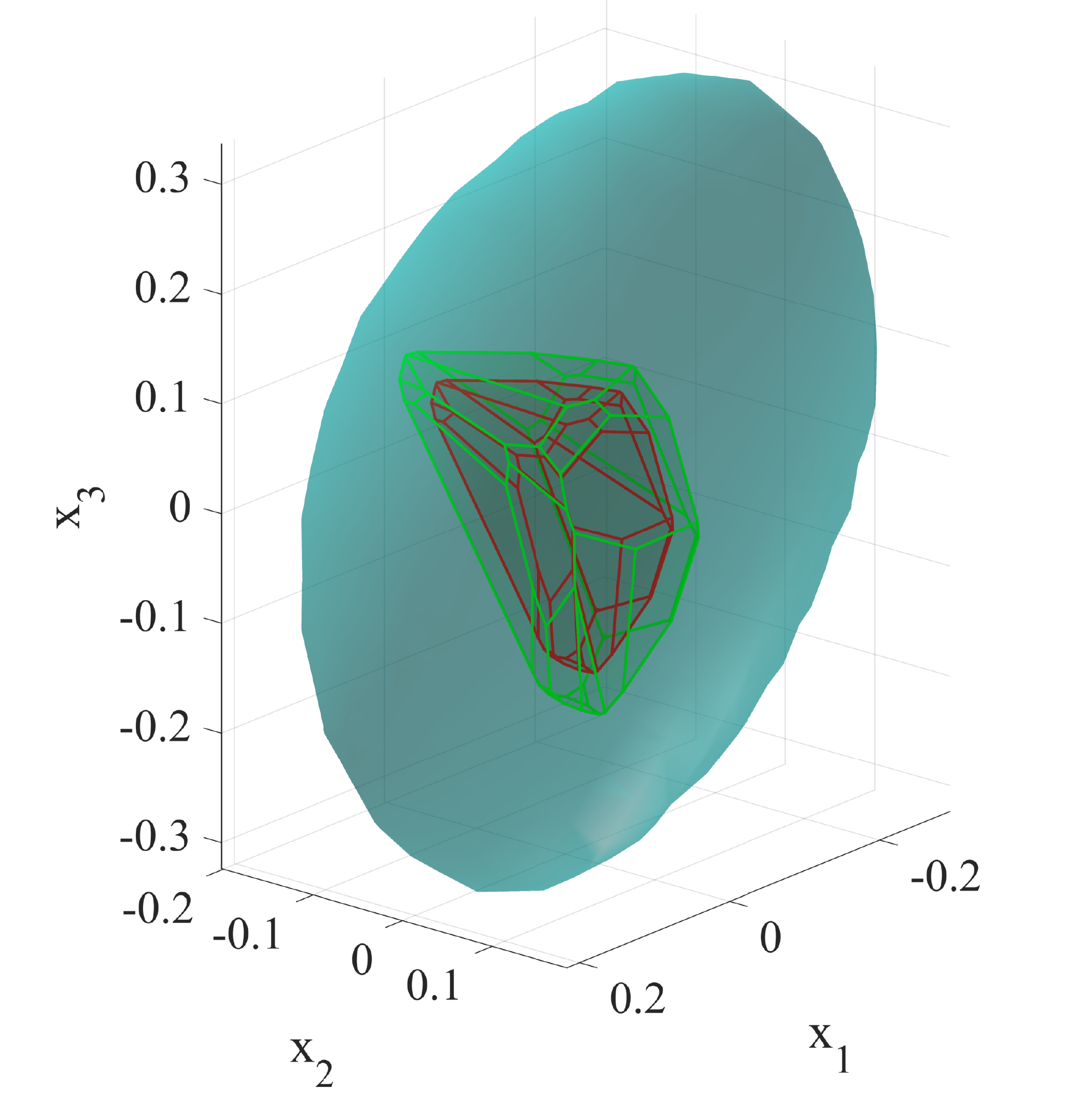}%
\label{f:samplepoly1000}}
\subfigure[LT-based (Lemma~1). $N_{LT}=52,044$]{\includegraphics[height=6cm]{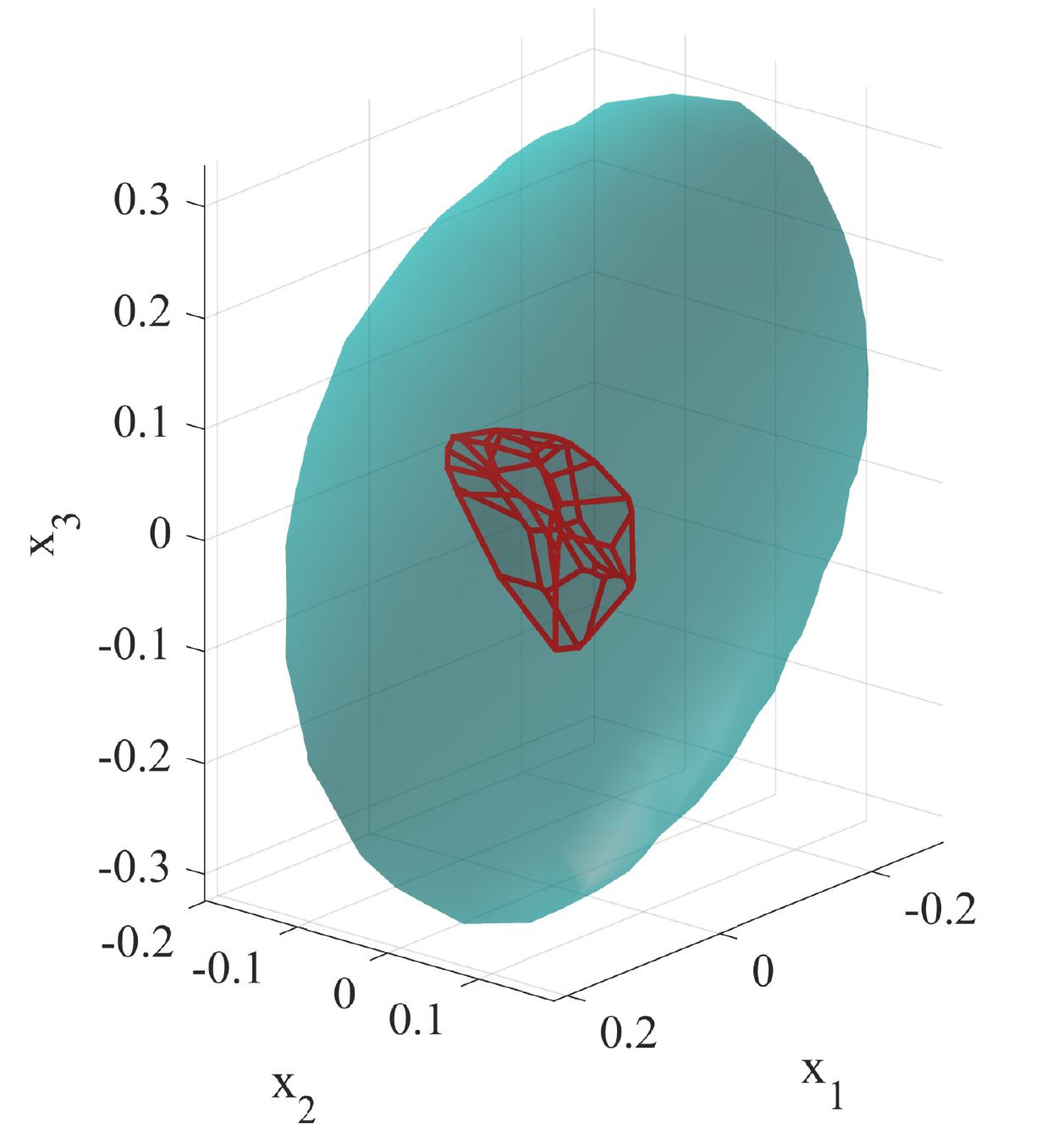}%
\label{f:samplepolyLT}}
\caption{(a-b) Probabilistic scaling approximations of the $\varepsilon$-CCS.
Scaling procedure applied to a sampled-polytope with $N_S=100$ (a) and $N_S=1,000$ (b). The initial sets are depicted in red, the scaled ones in green.
(c) Approximation obtained by direct application of Lemma \ref{lemma:scaling}. Note that, in this latter case, to plot the set without out-of-memory errors a pruning procedure \cite{herceg2013multi} of the $52,044$ linear inequalities was necessary. }
\label{f:samplepoly}
\end{figure*}
\begin{example}[Sample-based approximations]
To illustrate how the proposed scaling procedure works in practice in the case of sampled-polytope SAS, we revisit 
Example~\ref{ex:ill}.
To this end, a pre-fixed number $N_S$ of uncertainty samples were drawn, and the set inequalities 
\[
F(\tilde\q^{(j)})\theta\le  g(\tilde\q^{(j)}),\quad j\in\NATS{N_S},
\]
with $F(\q),g(\q)$ defined in \eqref{Fg_ex}, were constructed, leading to the candidate set $\Su_{N_S}$. Then, the corresponding Chebyshev center was computed, and Algorithm \ref{Algo1} was applied with $\varepsilon=0.05$, $\delta=10^{-6}$, leading to $N_\gamma=2,120$.

We note that, in this case, the solution of the optimization problem in \eqref{eq:opt-alg1} may be obtained by bisection on $\gamma$. Indeed, for given $\gamma$, checking if $\Su_{N_S}(\gamma)\subseteq\X(\q^{(i)})$
amounts to solving some simple linear programs.

Two different situations were considered: a case where the number of inequalities is rather small $N_S=100$, and a case where the complexity of the SAS is higher, i.e. $N_S=1,000$. The outcome procedure is illustrated in Figure~\ref{f:samplepoly}. We can observe that, for a small $N_S$ -- Fig.~\ref{f:samplepoly100}  -- the initial approximation is rather large (although it is contained in $\Xe$, we remark that we do not have any guarantee that this will happen). In this case, the probabilistic scaling returns $\gamma=0.8954$ which is less than one. This means that, in order to obtain a set fulfilling the desired probabilistic guarantees, we need to shrink it around its center. In the second case, for a larger number of sampled inequalities -- Fig.~\ref{f:samplepoly1000}  - the initial set (the red one) is much smaller, and the scaling procedure \textit{inflates} the set by returning a value of $\gamma$ greater than one, i.e. $\gamma=1.2389$. Note that choosing a larger number of samples for the computation of the initial set does not imply that the final set will be a better approximation of the $\varepsilon$-CCS.

Finally, we compare this approach to the scenario-like ones discussed in Subsection \ref{sec:rand}. To this end, we also draw the approximation obtained by directly applying the Learning Theory bound
\eqref{eq:Ntilde}.
Note that in this case, since $n_\theta=3$
and $n_\ell=4$, we need to take 
$N_{LT}=13,011$ samples, corresponding to $52,044$ linear inequalities. The resulting set is represented in Fig. \ref{f:samplepolyLT}. We point out that using this approximation i) the set is much more complex, since the number of involved inequalities is much larger, ii) the set is much smaller, hence providing a much more conservative approximation of the \CCS. Hence, the ensuing chance-constrained optimization problem will be computationally harder, and lead to a solution with a larger cost or even to an infeasible problem, in cases where the approximating set is too small.
\end{example}

\section{Candidate SAS: Norm-based SAS}
\label{sec:normSAS}
In this section, we propose a procedure in which the shape of the scalable SAS may be selected a-priori. This corresponds to situations where the designer wants to have full control in the final shape in terms of structure and complexity. 
The main idea is to define so-called \textit{norm-based SAS} of the form
\begin{equation}
\label{eq:normSAS}
\NSp(\gamma)\doteq\theta_c\oplus \gamma H \mathbb{B}_p^s
\end{equation}
where $\mathbb{B}_p^s$ is a $\ell_p$-ball in $\R^s$, $H\in\R^{n_\theta,s}$, with $s\ge n_\theta$, is a design matrix (not necessarily square), and $\gamma$ is the scaling parameter.
Note that when the matrix $H$ is square (i.e. $s=n_\theta$) and positive definite these sets belong to the class of $\ell_p$-norm based sets originally introduced in \cite{dabbene2010complexity}. In particular, in case of $\ell_2$ norm, the sets are ellipsoids. 
This particular choice is the one studied in \cite{CCTA2020}. Here, we extend this approach to a much more general family of sets, which encompasses for instance \textit{zonotopes}, obtained by letting  $p=\infty$ and $s\geq n_\theta$. Zonotopes have been  widely studied in geometry, and have found several applications in systems and control, in particular for problems of state estimation and robust Model Predictive Control, see e.g. \cite{alamo-zonotopes}.

\subsection{ Scaling factor computation for norm-bases SAS}
We recall that the scaling factor $\scale(\q)$ is defined as $0$ if $\theta_c\not \in \X(w)$ and as the largest value $\gamma$ for which $\NSp(\gamma) \subseteq \X(w)$ otherwise.
The following theorem, whose proof is reported in Appendix \ref{app3}, provides a direct and simple way to compute in closed form the scaling factor for a given candidate norm-based SAS.

\begin{theorem}[Scaling factor for norm-based SAS]\label{th2}
Given a norm-based SAS $\Sug$ as in \eqref{eq:normSAS}, 
and a realization $\q\in\Q$, the scaling factor $\scale(\q)$ can be computed as 
\[
\scale(\q)= \min_{\ell\in\NATS{n_\ell}} \; \gamma_\ell(w),
\]
with $\gamma_\ell(w)$, $\ell\in\NATS{n_\ell}$, given by
\begin{equation}
\label{gammaell}
 \gamma_\ell(w) = \bsis{ccl} 0 & \mbox{if } & \tau_\ell(w)  < 0, \\
\infty & \mbox{if} & \tau_\ell(w) \geq 0 \mbox{ and } \rho_\ell(w) =0,\\ 
\fracg{\tau_\ell(w)}{\rho_\ell(w)} &\mbox{if} &\tau_\ell(w) \geq 0  \mbox{ and } \rho_\ell(w) >0,\esis
\end{equation}
where 
$\tau_\ell(w) \doteq g_\ell(w) - f_\ell^T (w) \theta_c$ and 
$\rho_\ell(w) \doteq \| H^T f_\ell(w)  \|_{p^*} $, with $\|\cdot\|_p^*$ being the dual norm of $\|\cdot\|_p$.
\end{theorem}
Note that $\scale(\q)$ is equal to zero if and only if $\theta_c$ is not 
included in the interior of $\X(w)$.

\subsection{Construction of a candidate norm-based set}
\label{sec:normbasedSAS}

Similarly to Section~\ref{SAS-sampled-poly}, we first draw a fixed number $N_S$ of ``design" uncertainty samples
$\{\tilde\q^{(1)},\ldots,\tilde\q^{(N_S)}\},$ and construct an initial sampled approximation by introducing the following \textit{sampled-polytope SAS} $\Su_{N_S}$ as defined in $\eqref{eq:sampledSAS}$.
Again, we consider the Chebyshev center of $\Su_{N_S}$, or its analytical center as a possible center $\theta_c$ for our approach. 

Given $\Su_{N_S}$, $s\geq n_{\theta}$ and $p\in\{1,2,\infty\}$, the objective is to compute the largest set $\theta_c\oplus H \mathbb{B}^s_p$ included in $\Su_{N_S}$. 
To this end, we assume that we have a function $\volp(H)$ that provides a measure of the size of $H\mathbb{B}^s_p$. That is, larger values of  $\volp(H)$ are obtained for increasing sizes of $H\mathbb{B}^s_p$.

\begin{remark}[On the volume function]~\\
The function $\volp(H)$ may be seen as a generalization of the classical concept of Lebesgue volume of the set $\Su_{N_S}$.
Indeed, when $H$ is a square positive definite matrix, some possibilities are $\volp(H)=\log\,\det(H)$ -- which is directly proportional to the classical volume definition, or $\volp(H)=\rm{tr}\,H$ -- which for $p=2$ becomes the well known sum of ellipsoid semiaxes (see \cite{dabbene2015randomized} and \cite[Chapter 8]{Boyd04}). 
These measures can be easily generalized to non square matrices. It suffices to compute the singular value decomposition. If $H=U\Sigma V^T$, we could use the measures $\volp(H)=\rm{tr}\,\Sigma$ or $\volp(H)=\log\,\det(\Sigma)$. \\
For non square matrices $H$, specific results for particular values of $p$ are known.  For example, we remind that if $p=\infty$ and $H\in \R^{n_\theta\times s}$, $s\geq n_\theta$, then $\theta_c\oplus H\mathbb{B}^s_\infty$ is a zonotope. Then, if we denote as \textit{generator} each of the columns of $H$,
the volume of a zonotope  can be computed by means of a sum of terms (one for each different way of selecting $n_{\theta}$ generators out of the $s$ generators of $H$); see \cite{alamo2005guaranteed}, \cite{gover2010determinants}. Another possible measure of the size of a zonotope $\theta_c\oplus H\mathbb{B}^s_\infty$ is the Frobenious norm of $H$ \cite{alamo2005guaranteed}. 

\end{remark}

Given an initial design set $\Su_{N_S}$, we elect as our candidate Scalable SAS the  largest ``volume" norm-based SAS contained in $\Su_{N_S}$. Formally, this rewrites as the following optimization problem 
\begin{align*} 
&\max\limits_{\theta_c,H} ~\volp(H) \\
&\text{subject to }  \theta_c\oplus H\mathbb{B}_p^s \subseteq 
\Su_{N_S}
%\bigcap\limits_{j=1}^{N_S} \X(w^{(j)}).
\end{align*}
As it has been shown, this problem is equivalent to 
\begin{eqnarray*} \min\limits_{\theta_c,H} && -\volp(H) \\
\text{s.t.} && f_\ell^T(\tilde\q^{(j)}) \theta_c +  \| H^T f_\ell(w^{(j)})  \|_{p^*}-g_\ell(w^{(j)}) \leq 0,\\
&& \qquad\qquad\qquad\ell\in\NATS{n_\ell}, \; j\in\NATS{N_S}, 
\end{eqnarray*}
where we have replaced the maximization of $\volp(H)$ with the minimization of -$\volp(H)$.% Moreover, we assume that $p^*=2$ if $p=2$, $q=1$ if $p=\infty$ and $q=\infty$ if $p=1$.

We notice that the constraints are convex on the decision variables; also, the functional to minimize is convex under particular assumptions. For example when $H$ is assumed to be square and positive definite and $\volp(H)=\log\det (H)$. 
For non square matrices, the constraints remain convex, but the convexity of the functional to be minimized is often lost. In this case, local optimization algorithms should be employed to obtain a possibly sub-optimal solution.
\begin{figure*}[!ht]
\centering
\subfigure[$\gamma=0.9701$]{\includegraphics[height=4.8cm]{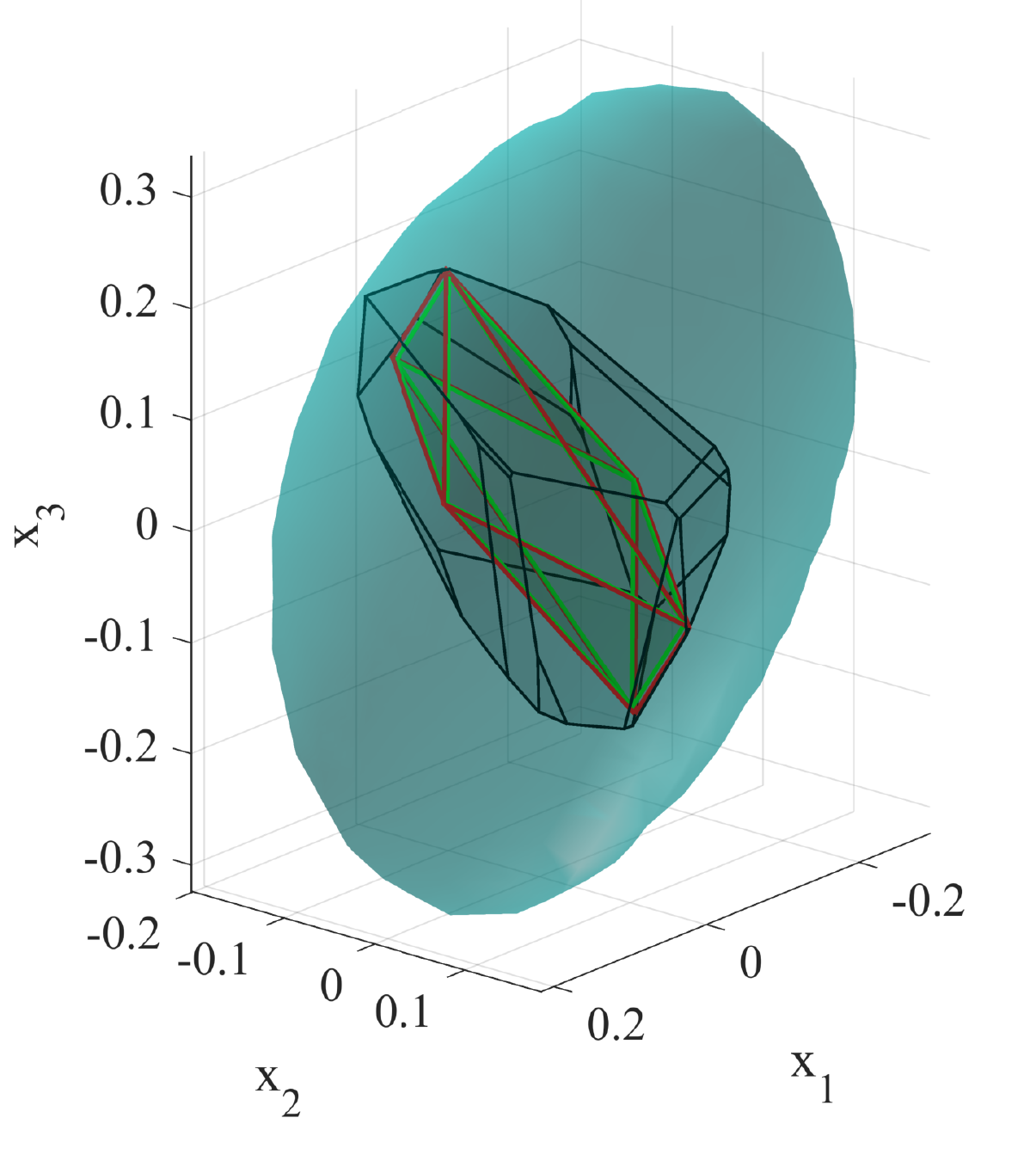}%
\label{f:diam100}}
\subfigure[$\gamma=1.5995$]{\includegraphics[height=4.8cm]{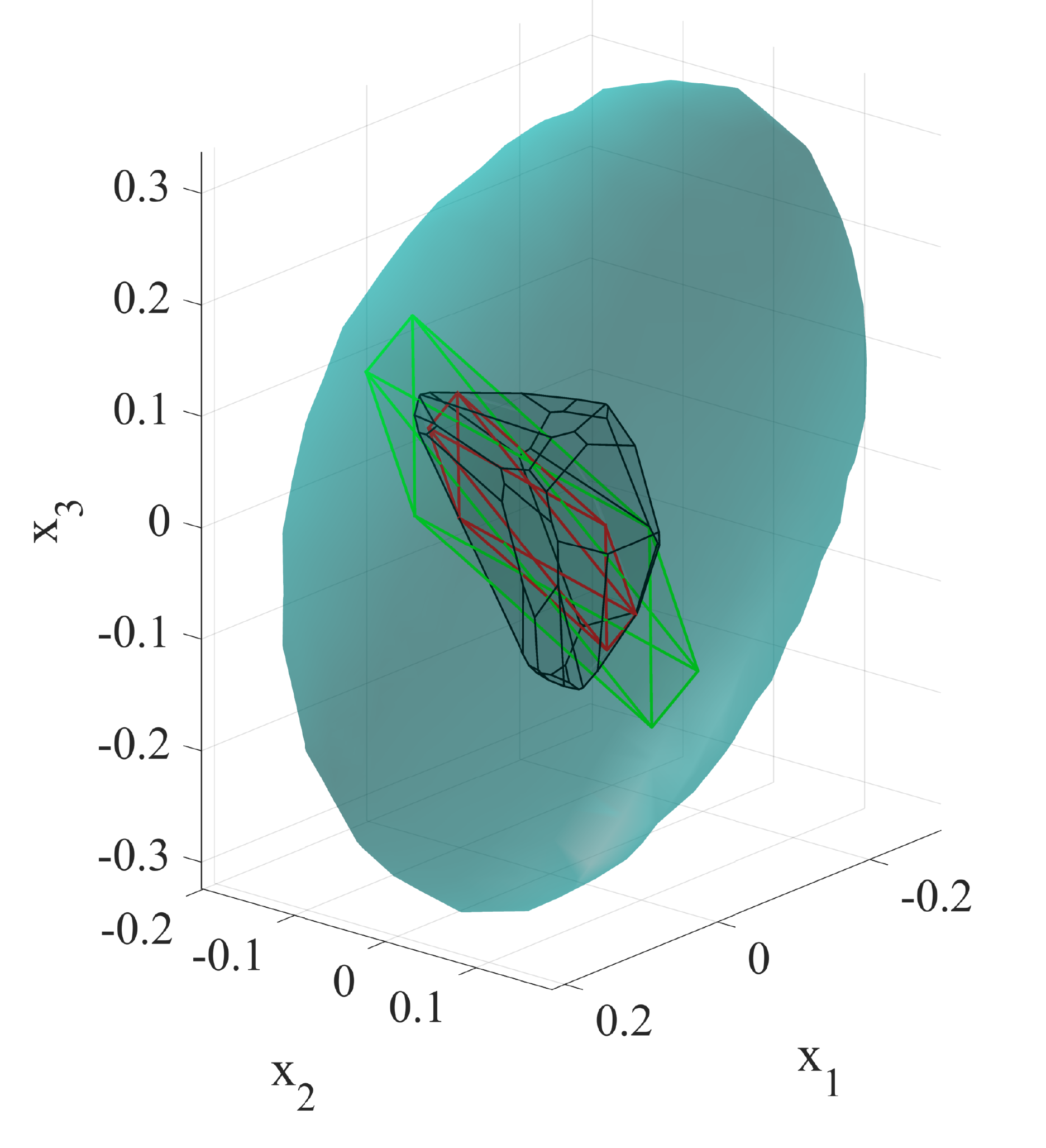}%
\label{f:diam1000}}
\subfigure[$\gamma=0.9696$]{\includegraphics[height=4.8cm]{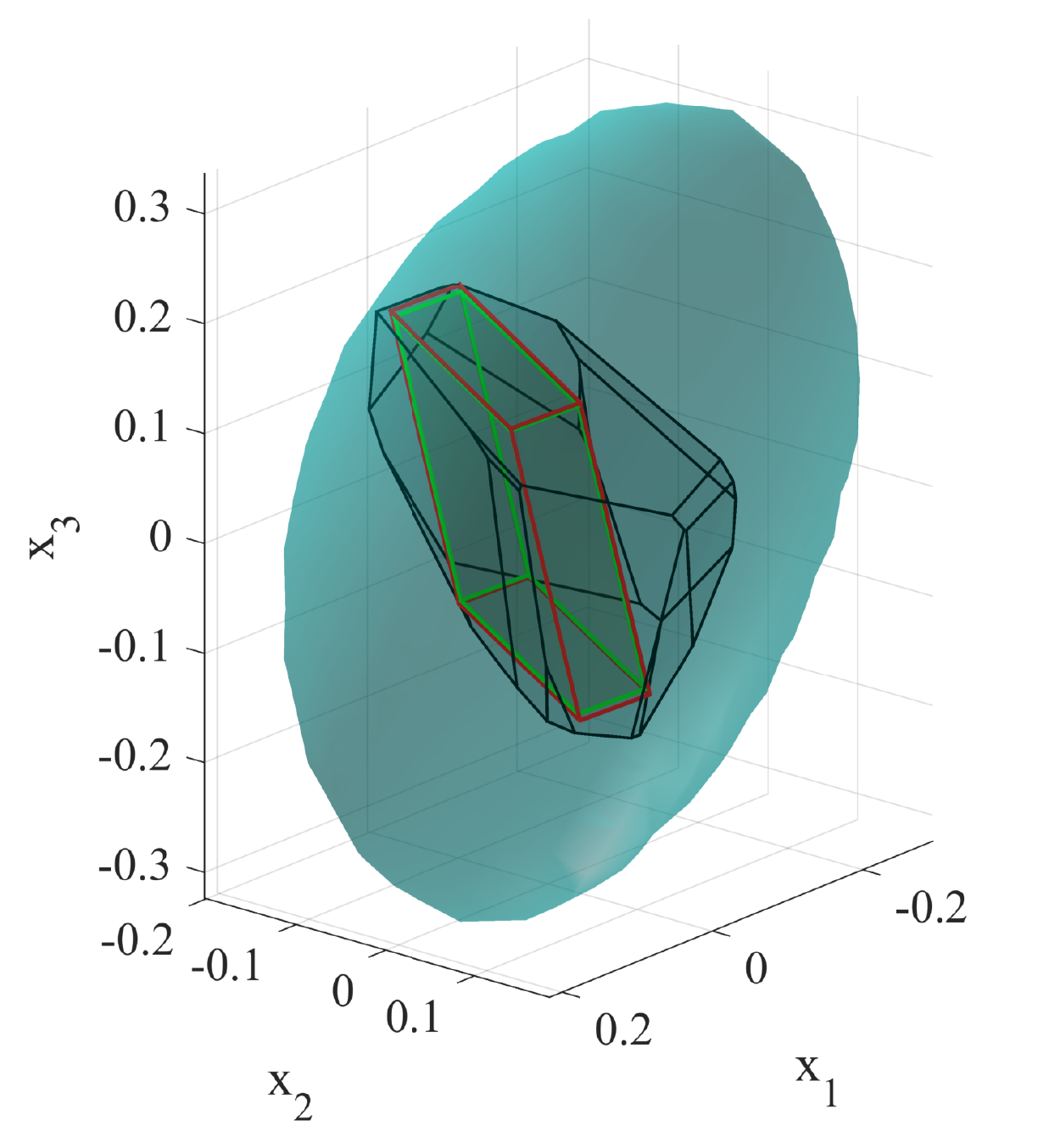}%
\label{f:hyper100}}
\subfigure[$\gamma=1.5736$]{\includegraphics[height=4.8cm]{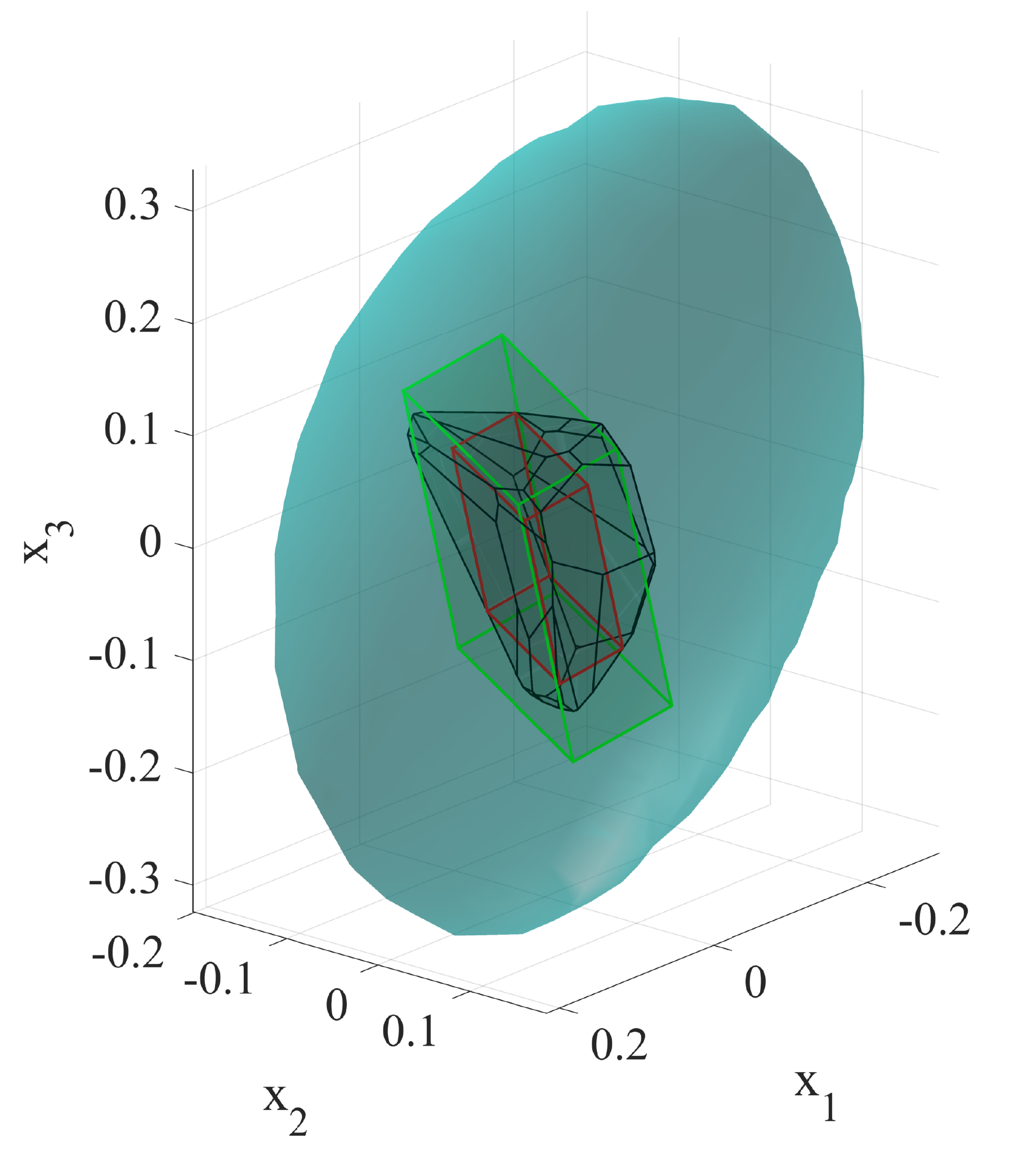}%
\label{f:hyper1000}}
\caption{Scaling procedure applied to (a) $\Su_1$-SAS with $N_S=100$, (b) $\Su_1$-SAS with $N_S=1,000$ (b),  $\Su_\infty$-SAS with $N_S=100$ (c), and $\ell_\infty$-poly with $N_S=1,000$ (d).
The initial set is depicted in red, the final one in green. The sampled design polytope $\Su_{N_S}$ is represented in black. }
\label{f:diam-hyper}
\end{figure*}

\medskip

\begin{example}[Norm-based SAS]
We revisit again Example~\ref{ex:ill} to show the use of norm-based SAS. We note that, in this case, the designer can control the approximation outcome by acting upon the number of design samples $N_S$ used for constructing the set $\Su_{N_S}$. In Figure~\ref{f:diam-hyper} we report two different norm-based SAS, respectively with $p=1$ and $p=\infty$, and for each of them we consider two different values of $N_S$, respectively $N_S=100$ and $N_S=1,000$. Similarly to what observed for the sampled-polys, we see that for larger $N_S$, the ensuing initial set becomes smaller. Consequently, we have an inflating process for small $N_S$ and a shrinkage one for large $N_S$ %\VIC{The scaling phase will inflate the SAS with small $N_S$ and shrink those with large $N_S$}.
However, we observe that in this case, the final number of inequalities is independent on $N_S$, being equal to $3n_\theta+1=10$ 
for\,\,$\Su_1$ and  $2n_\theta$ for $\Su_\infty$.
\end{example}

\medskip

\subsubsection{Relaxed computation}
It is worth remarking that that the minimization problem of the previous subsection might be infeasible. In order to guarantee the feasibility of the problem, a soft-constrained optimization problem is proposed. With a relaxed formulation, $\theta_c$ is not guaranteed to satisfy all the sampled constraints.  However $\theta_c\in \Su_{N_S}$ is not necessary to obtain an $\varepsilon$-CSS (in many practical applications, every element of $\Theta$ has a non zero probability of violation and $\Su_{N_S}$ is empty with non-zero probability). Moreover, a relaxed formulation is necessary to address problems in which there is no element of $\Theta$ with probability of violation equal to zero (or significantly smaller than $\varepsilon$). Not considering the possibility of violations is an issue especially when $N_S$ is large, because the probability of obtaining an empty sampled set $\Su_{N_S}$ grows with the number of samples $N_S$. 

Given $\xi>0$ the relaxed optimization problem is 
\begin{align} 
& \min\limits_{\theta_c,H,
\tau_1,\ldots,\tau_{N_S}}~ -\volp(H) + \xi \Sum{j=1}{N_S} \max\{\tau_j,0\} \label{prob:relaxed}\\
& \text{s.t. }\; 
f_\ell^T(w^{(j)}) \theta_c + \| H^T f_\ell(w^{(j)})  \|_{p^*}-g_\ell(w^{(j)}) \leq \tau_j,\nonumber\\
& \qquad\qquad\qquad\ell\in\NATS{n_\ell}, \; j\in\NATS{N_S}. \nonumber
\end{align}
The parameter $\xi$ serves to provide an appropriate trade off between satisfaction of the sampled constraints and the size of the obtained region. A possibility to choose $\xi$ would be to choose it in such a way that the fraction of violations $n_{viol}/N_S$ (where $n_{viol}$ is the number of elements $\tau_j$ larger than zero) is smaller than $\varepsilon/2$. 

\section{Numerical example: Probabilistic set membership estimation}
\label{sec:num_ex}
We now present a numerical example in which the results of the paper are applied to the probabilistic set membership estimation problem, introduced in subSection \ref{sub:sec:Second:Motivating:Set:membeship}. We consider the universal approximation functions given by Gaussian radial basis function networks (RBFN) \cite{buhmann2000radial}. 

Given the nodes $[x_1,x_2, \ldots, x_M]$ and the variance parameter $c$, the corresponding Gaussian radial basis 
function network is defined as
$$\RBFN(x,\theta)= \theta^T \varphi(x), $$
where $\theta=\bmat{ccc} \theta_1 & \hdots &\theta_M\emat^T$ represents the weights and $$\varphi(x)= \bmat{ccc} \exp\left(\frac{-\|x-x_1\|^2}{c}\right) & \ldots & \exp\left(\frac{-\|x-x_M\|^2}{c}\right)\emat^T$$ is the regressor function. 
Given $\delta\in(0,1)$ and $\varepsilon\in (0,1)$,  the objective is to obtain, with probability no smaller than $1-\delta$, an inner approximation of the probabilistic feasible parameter set $\FPS_\varepsilon$, which is the set of parameters $\theta\in \R^M$ that satisfies 
\begin{equation}\label{eq:fps:set-membership} 
 \Pq \{ | y-\theta^T \varphi(x)| \leq \rho \} \geq 1-\varepsilon,
 \end{equation}
where $x$ is a random scalar with uniform distribution in $[-5,5]$ and 
$$y = \sin(3 x) + \sigma,$$
where $\sigma$ is a random scalar with a normal distribution with mean $5$ and variance 1.

%To this end, we first tune $\theta$ to estimate the probabilistic set membership $\FPS_\varepsilon$, so that, if $\theta \in \FPS_\varepsilon$, then, with confidence $1-\delta$:

%\begin{equation}
%\label{eq:fps:set-membership}
%\Pq \{ |y-\theta^T \varphi(x)| \leq \rho \} \geq 1-\varepsilon.
%\end{equation}

%We see that this expression is the same as Equation \ref{equ:pFPS}. 
We use the procedure detailed in Sections \ref{sec:scaling}, \ref{SAS-sampled-poly} and \ref{sec:normSAS} to obtain an SAS of $\FPS_\varepsilon$. We have taken a grid of $M=20$ points in the interval $[-5,5]$ to serve as nodes for the RBFN, and a variance parameter of $c=0.15$. We have taken $N_S=350$ random samples  $w=(x,y)$ to compute the initial geometry, which has been chosen to be  an $\ell_\infty$ norm-based SAS of dimension 20 with a relaxation parameter of $\xi=1$ (see \eqref{prob:relaxed}). The chosen initial geometry is $\theta_c\oplus H\mathbb{B}^{20}_\infty$, where $H$ is constrained to be a diagonal matrix. 

\begin{figure}[!ht]
\centering
%\def\svgwidth{\columnwidth}
%\import{figures/}{int-pred-fps2.pdf_tex}
\includegraphics[width=1\columnwidth]{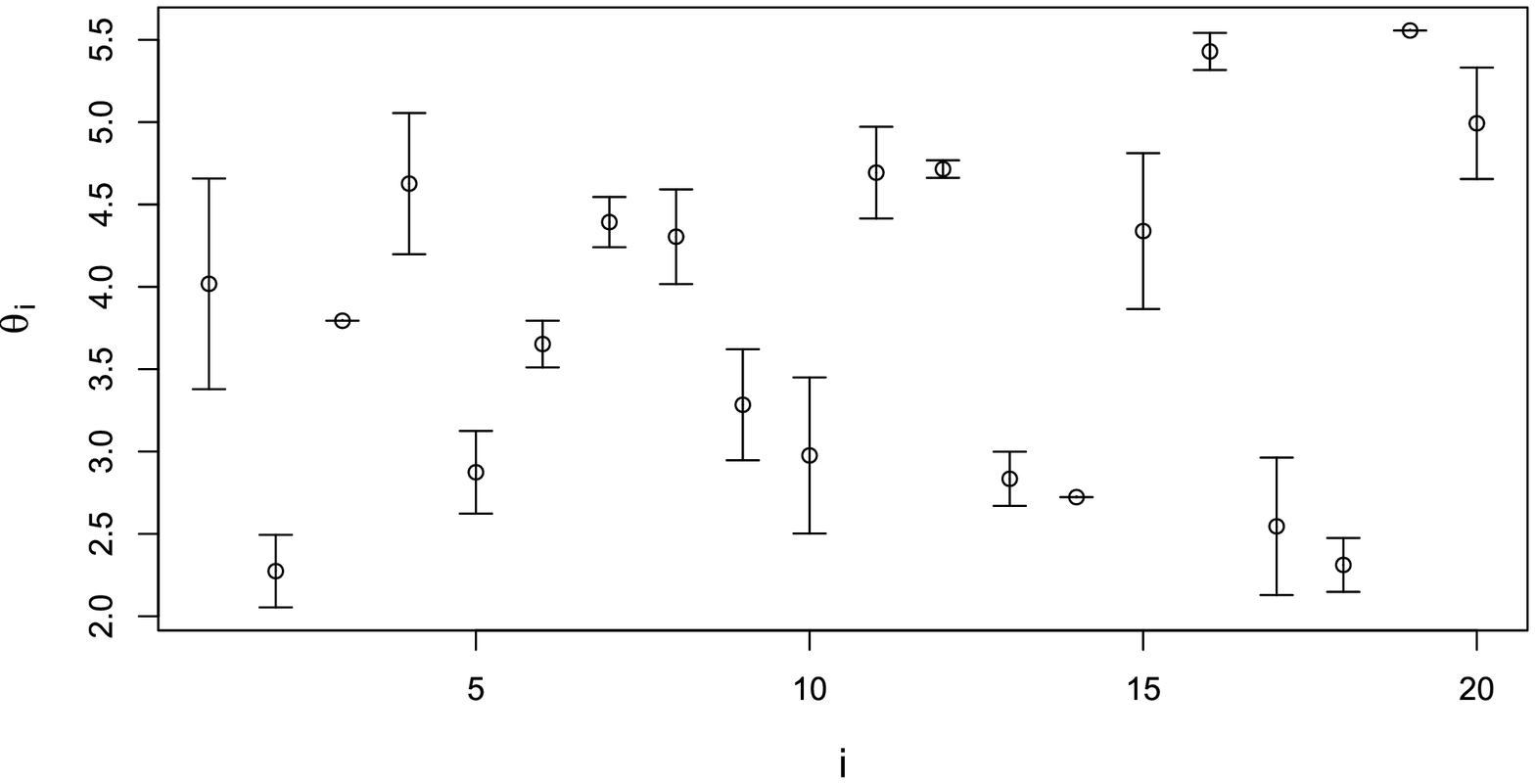}%
\caption{Representation of the extreme values $\theta^+$ and $\theta^-$ and the central value $\theta_c$ of the $\FPS_\varepsilon^{\delta}$.}
\label{f:ipe2}
\end{figure}

When the initial geometry is obtained, we scale it around its center by means of probabilistic scaling with Algorithm \ref{alg:scaling}. The number of samples required for the scaling phase to achieve $\varepsilon=0.05$ and $\delta=10^{-6}$ is $N_\gamma=2065$ and the resulting scaling factor is $\gamma=0.3803$. The scaled geometry $\theta_c\oplus\gamma H\mathbb{B}^{20}_\infty$ is, with a probability no smaller than $1-\delta$, an inner approximation of $\FPS_\varepsilon$ which we will refer to as $\FPS_\varepsilon^{\delta}$. Since it is a transformation of an $\ell_\infty$ norm ball with a diagonal matrix $H$, we can write it as
\vskip -2mm
$$\FPS_\varepsilon^{\delta} = \{ \theta : \theta^- \leq \theta \leq \theta^+ \},$$
\vskip -2mm
where the extreme values $\theta^-, \theta^+\in\R^{20}$ are represented in Figure \ref{f:ipe2} \cite{plotrix}, along with the central value $\theta_c\in \R^{20}$.

\vskip 1mm

\begin{figure}[!ht]
\centering
%\def\svgwidth{\columnwidth}
%\import{figures/}{int-pred-example4.pdf_tex}
\includegraphics[width=1\columnwidth]{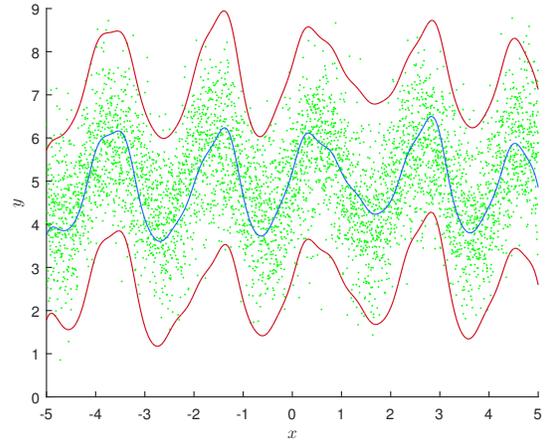}%
\caption{Real values of $y$ vs central estimation (blue) and interval prediction bounds (red).}
\label{f:ipe1}
\end{figure}

Once the $\FPS_\varepsilon^{\delta}$ has been computed, we can use its center $\theta_c$ to make the point estimation $y \approx \theta_c^T \varphi(x)$. We can also obtain probabilistic upper and lower bounds of $y$ by means of equation \eqref{eq:fps:set-membership}. That is, every point in $\FPS_\varepsilon^{\delta}$ satisfies, with confidence $1-\delta$:
\begin{equation}
\begin{aligned}
&\Pq \{ y \leq \theta^T \varphi(x) + \rho \} \geq 1-\varepsilon, \\
&\Pq \{ y \geq \theta^T \varphi(x) - \rho \} \geq 1-\varepsilon.
\end{aligned}
\end{equation}
We notice that the tightest probabilistic bounds are obtained with $\theta^+$ for the lower bound and $\theta^-$ for the upper one. That is, we finally obtain that,  with confidence $1-\delta$:
\begin{equation}
\begin{aligned}
&\Pq \{ y \leq {\theta^-}^T \varphi(x) + \rho \} \geq 1-\varepsilon, \\
&\Pq \{ y \geq {\theta^+}^T \varphi(x) - \rho \} \geq 1-\varepsilon. 
\end{aligned}
\end{equation}
Figure \ref{f:ipe1} shows the results of both the point estimation and the probabilistic interval estimation.

\section{Conclusions, extensions, and future directions}
\label{sec:binary}
In this paper, we proposed a general approach to construct probabilistically guaranteed inner approximations of the chance-constraint set $\Xe$. The approach is very general and flexible.

First, we remark that the proposed scaling approach is not limited to 
sets defined by linear inequalities, but immediately extends to more general sets.
Indeed, we may consider a generic binary performance function $ \phi: \Theta \times \Q \to\{0,\,1\}$ defined as \footnote{Clearly, this formulation encompasses the setup discussed, obtained by simply setting 
$
\phi(\theta,\q)=
\left\{
\begin{array}{ll}
0 &\text{if $F(\q)\theta\le  g(\q)$}\\
1 &\text{otherwise.}
\end{array}\right.
$}
\begin{equation}
\label{spec}
\phi(\theta,q)=
\left\{
\begin{array}{ll}
0 &\text{if $\theta$ meets design specifications for $\q$}\\
1 &\text{otherwise.}
\end{array}\right.
\end{equation}
In this case, the violation probability may be written as $\viol(\theta)\doteq \Pr_{\Q}\,\{\,\gi(\theta,\q)=1\,\}=\EE(\theta)$,
and we can still define the set $\Xe$ as in \eqref{Xe}. Then, given an initial SAS candidate, Algorithm 1 still provides a valid approximation.
However, it should be remarked that, even if we choose a ``nice" SAS as those previously introduced, the nonconvexity of $\phi$
will most probably render step 4 of the algorithm intractable.
To further elaborate on this point, let us focus on the case when the design specification may be expressed as a (nonlinear) inequality of the form
\[
\gi(\theta,q)\le 0.
\]
Then, step 4 consist in solving the following nonconvex optimization problem
\begin{align}
    \gamma_i \doteq &\arg\max \gamma \\
    &\text{s.t.}\quad  \Sug \subseteq \X(\q^{(i)})  =\Bigl\{\theta\in\Theta\;|\;
\gi(\theta,\q^{(i)})\le 0
\Bigr\}.\nonumber
\end{align}
We note that this is general a possibly hard problem. However, there are cases when this problem is still solvable.
For instance, whenever $\gi(\theta,q)$ is a convex function of $\theta$ for fixed $\q$ and the set $\Su$ is also convex, 
the above optimization problem may be formulated as a convex program by application of  Finsler lemma.
We remark that, in such situations, the approach proposed here is still completely viable, since all the derivations continue to hold.

Second, we remark that the paper open the way to the design of other families of Scaling SAS. For instance, we are currently 
working on using the family of sets defined in the form of polynomial superlevel sets (PSS) proposed in \cite{Dabbene-PSS}.
\appendix

\section{Appendix}
\subsection{Proof of Lemma 1}
\label{app:proof-Lemma1}

To prove the lemma, we first recall the following definition from  \cite{alamo2009randomized}.

\begin{definition}[$(\alpha,k)$-Boolean Function] The function $h: \Theta\times\Q\to \mathbb{R}$
is an $(\alpha,k)$-Boolean function if for fixed $\q$  it can be written as an expression consisting of Boolean operators involving $k$ polynomials
$
p_1(\theta),p_2(\theta),\ldots,p_k(\theta),
$
in the components  $\theta_i$, $i\in\NATS{n_\theta}$ and the degree with respect to  $\theta_i$ of all these polynomials is no larger than $\alpha$.
\end{definition}
Let us now define the binary functions 
\[
h_\ell(\theta,\q) \doteq \left\{ \begin{array}{rl} 0 & \mbox{ if } f_\ell(\q)\theta \le g_\ell(\q) \\ 1 & \mbox{ otherwise}  \end{array}\right.,\; \ell\in\NATS{n_\ell}. 
\]
Introducing the function $h(\theta,\q) \doteq \max\limits_{\ell=1,\ldots,n_\ell} h_\ell(\theta,\q),$
we see that the violation probability can be alternatively written as $\viol(\theta)\doteq \Pr_{\Q}\,\{\,h(\theta,\q)=1\,\}.$
The proof immediately follows by observing that
$h(\theta,\q)$ is an $(1,n_\ell)$-Boolean function, since it can be expressed as a function of $n_\ell$ Boolean functions, each of them involving a polynomial of degree 1. 
Indeed, it is proven in \cite[Theorem 8]{alamo2009randomized}, that, if $h: \Theta\times\Q\to \mathbb{R}$
is an $(\alpha,k)$-Boolean function then, 
for $\varepsilon\in(0,0.14)$,  with probability greater than $1-\delta$ we have 
$
\Pr_{\Q}\,\{\,h(\theta,\q)=1\,\}\le \varepsilon
$
if $N$ is chosen such that
\begin{multline*}
N\ge \frac{4.1}{\varepsilon}\Big(\ln \frac{21.64}{\delta}+4.39n_\theta\,\log_{2}\Big(\frac{8e\alpha k}{\varepsilon}\Big)\Big).
\end{multline*}

\subsection{Property \ref{prop:scale:sets}}\label{app2}

\begin{property}\label{prop:scale:sets}
Given  $\varepsilon\in(0,1)$,  $\delta\in (0,1)$, and  $0\le r \leq N$, let $N$ be such that
$\Bin(r;N,\varepsilon)\leq \delta$.
Draw $N$ iid sample-sets $ \{ \X^{(1)}, \X^{(2)}, \ldots, \X^{(N)} \} $
from a distribution $\Pr_{\X}$. For $i\in\NATS{N}$,  let
$ \gamma_i \doteq\scale(\X^{(i)})$,  with $\scale(\cdot)$ as in Definition~\ref{def-scaling},
and suppose that $  \bar\gamma =    \gamma_{1+r:N} >0$. Then, with probability no smaller than $1-\delta$, 
it  holds that $\Pr_{\X} \{\theta_c \oplus \bar\gamma \Su \not \subseteq \X\} \leq \varepsilon$.
\end{property}

\noindent
\noindent\textit{Proof:} 
It has been proven in \cite{CalafioreSIAM10,Campi11} that if one discards no more than $r$ constraints on a convex problem with $N$ random constraints,  then the probability of violating the constraints with the solution obtained from the random convex problem is no larger than $\varepsilon\in(0,1)$, with probability no smaller than $1-\delta$, where
$$ \delta =\conv{d+r-1}{d-1} \Sum{i=0}{d+r-1}\conv{N}{i}\varepsilon^i(1-\varepsilon)^{N-i},$$
and $d$ is the number of decision variables. 
We apply this result to the following optimization problem
\begin{equation}
 \label{minEEE}
 \max \limits_{\gamma}  \gamma \text{ subject to }
  \theta_c \oplus \gamma \Su \subseteq \X^{(i)}, \;\; i\in\NATS{N}.\nonumber
\end{equation}
From Definition~\ref{def-scaling}, we could rewrite this optimization problem as
\[ \max \limits_{\gamma} \gamma  \text{ subject to }  \gamma \leq\scale(\X^{(i)}),\; i\in\NATS{N}.
\]
We first notice that the problem under consideration is convex and has a unique scalar decision variable $\gamma$. That is, $d=1$. Also, the non-degeneracy and uniqueness assumption required in the application of the results of \cite{CalafioreSIAM10} and \cite{Campi11} are satisfied. Hence, if we allow $r$ violations in the above minimization problem, we have that with probability no smaller than $1-\delta$, where
\[
 \delta = \conv{r}{0} \Sum{i=0}{r}\conv{N}{i}\varepsilon^i(1-\varepsilon)^{N-i}=\Bin(r;N,\varepsilon),
 \]
the solution $\bar\gamma$ of problem (\ref{minEEE}) satisfies
$ \Pr_{\X} \{ \bar\gamma >\scale(\X) \} \leq \varepsilon.$
We conclude from this, and Definition~\ref{def-scaling}, that with probability no smaller than $1-\delta$,
$$  \Pr_{\X} \{ \theta_c \oplus \bar\gamma \Su \not\subseteq \X\} \leq \varepsilon. $$ 
Finally, note that the optimization problem under consideration can be solved directly by ordering the values $\gamma_i=\scale(\X^{(i)})$. It is clear that if $r\geq 0$ violations are allowed, then the optimal value for $\gamma$ is $\bar\gamma=\gamma_{r+1:N}$.
\hfill\qedsymbol

\subsection{Proof of Theorem \ref{th2}}\label{app3}
Note that, by definition, the condition
$\theta_c\oplus \gamma H\mathbb{B}^s_p \subseteq \X(w)$
is equivalent to 
\[
 \max\limits_{z\in \mathbb{B}^s_p}  f_\ell^T(w)(\theta_c + \gamma Hz)-g_\ell(w) \leq 0, \; \ell\in\NATS{n_\ell}.
\]
Equivalently, from the dual norm definition, we have
\[
f_\ell^T(w) \theta_c + \gamma \| H^T f_\ell(w)  \|_{p^*}-g_\ell(w) \leq 0, \; \ell\in\NATS{n_\ell}. 
\]
Denote by $\gamma_\ell$ the scaling factor $\gamma_\ell$ corresponding to the $\ell$-th constraint
\[
f_\ell^T(w) \theta_c + \gamma_\ell \| H^T f_\ell(w)  \|_{p^*}-g_\ell(w) \leq 0.
\]
With the notation introduced in the Lemma, this constraint rewrites as
\[
\gamma_\ell \rho_\ell(w) \leq \tau_\ell(w).
\]
The result follows noting that the
corresponding scaling factor $\gamma_\ell(\q)$ can be computed as 
\[
\gamma_\ell(\q)=\max_{\gamma_\ell \rho_\ell(w) \leq \tau_\ell(w)}\gamma_\ell,
\]
and that the value for $\gamma(\q)$ is   obtained   from   the   most restrictive one.
\qedsymbol\\

%%%%%%%%%%%%%%%%%%%%%%%%%%%%%%%%%%%%%%%%%%%%%%%%%%%%%%%
%\end{comment}
\bibliographystyle{IEEEtran}
%\small
\bibliography{BIBLIO}             % bib file to %

\end{document}